\documentclass[12pt]{article}
\usepackage{amsmath}
\usepackage{graphicx}
\usepackage{enumerate}
\usepackage{multirow}
\usepackage{natbib} 
\usepackage{url} 

\usepackage{float} 

\usepackage{verbatim}

\newcommand{\blind}{0}

\newcommand{\bc}{\mathbf{c}}

\newcommand{\bbf}{\mathbf{f}}
\newcommand{\bg}{\mathbf{g}}
\newcommand{\bh}{\mathbf{h}}

\newcommand{\bu}{\mathbf{u}}
\newcommand{\bv}{\mathbf{v}}

\newcommand{\by}{\mathbf{y}}
\newcommand{\bz}{\mathbf{z}}

\newcommand{\bP}{\mathbf{P}}
\newcommand{\bQ}{\mathbf{Q}}

\newcommand{\bW}{\mathbf{W}}
\newcommand{\bX}{\mathbf{X}}

\newcommand{\bbeta}{\boldsymbol{\beta}}

\newcommand{\bgamma}{\boldsymbol{\gamma}}

\begin{document}

\def\spacingset#1{\renewcommand{\baselinestretch}%
{#1}\small\normalsize} \spacingset{1}


\if0\blind
{
  \title{Hybrid Smoothing for Anomaly Detection in Time Series}
  \author{Matthew Hofkes \\
    Department of Statistics, Colorado School of Mines \\
    Douglas Nychka \\
    Department of Statistics, Colorado School of Mines \\
    Tzahi Cath \\
    Department of Environmental Engineering, Colorado School of Mines \\
    Amanda Hering \\
    Department of Statistics, Baylor University \\
    Craig McGonagill \\
    Denver Water, Denver, Colorado \\
    }
  \maketitle

} \fi

\if1\blind
{
  \bigskip
  \bigskip
  \bigskip
  \begin{center}
    {\LARGE\bbf Title}
\end{center}
  \medskip
} \fi

\vspace{-2em}
\begin{abstract}
Many industrial and engineering processes monitored as times series have smooth trends that indicate normal behavior and occasionally anomalous patterns that can indicate a problem.  This kind of behavior can be modeled by a smooth trend, such as a spline or Gaussian process, and a disruption based on a sparser representation.  Our approach is to expand the process signal into two sets of basis functions: one set uses $L_2$ penalties on the coefficients, and the other set uses $L_1$ penalties to control sparsity. From a frequentist perspective, this results in a hybrid smoother that combines cubic smoothing splines and the LASSO. As a Bayesian hierarchical model (BHM), this is equivalent to priors giving a Gaussian process and a Laplace distribution for anomaly coefficients. For the hybrid smoother, we propose two new ways of determining the penalty parameters that use effective degrees of freedom and contrast this with the BHM that uses loosely informative inverse gamma priors.  Several reformulations are used to make sampling the  BHM  posterior more efficient, including some novel features in orthogonalizing and regularizing the model basis functions.  This methodology is motivated by a substantive application, offline monitoring of a water treatment  process for municipal water filtration.  We also test the robustness of these methods with a Monte Carlo study designed to inspect a range trended time series under an array of conditions and compare this new approach to multiple existing modern methods. Both the hybrid smoother and the full BHM give comparable results with small false positive and false negative rates.  Besides being successful in the water treatment application, this work can be easily extended to other Gaussian process models and other features that represent process disruptions in offline data.   \end{abstract}

\noindent%
\vfill

\newpage
\spacingset{1} 
\section{Introduction}
\label{sec:intro}

Many industrial and engineering processes that are monitored as times series have smooth trends that indicate normal behavior and occasionally  anomalous patterns that can indicate a problem. In this work, we combine a flexible model for a smooth curve with the potential to include discontinuous interruptions.  Although there are many possibilities for introducing anomalous behavior, this work is motivated by a specific problem in water filtration.   Here, normal operation appears to be well described by a Gaussian process (GP), and the anomalous intervals exhibit abrupt discontinuities added to this GP baseline.  From a frequentist point of view, we characterize a GP model as involving a basis function expansion with quadratic penalties on the basis coefficients.  On the other hand, the anomalous signal tends to  have discontinuities that are limited in number and so suggests a basis representation where many coefficients are zero. This sparse representation can be captured by $L_1$ penalties on the basis coefficients.  Thus, we are led to a hybrid process signal that is a sum of two components, each based on a separate set of basis functions. One set uses $L_2$ penalties to identify the trend while the other set uses $L_1$ penalties to identify disturbances.   The development of the LASSO and related models  $L_1$ penalties have been very successful in building sparsity into a set of parameters in an efficient and practical manner.  However, the combination of $L_2$ and $L_1$ models for the basis coefficients to identify a mixed signal has not been exploited.  

In the context of monitoring a process, one often has a record of past system performance and some prior knowledge of the normal and anomalous behavior. Thus, it is natural to approach this problem also from a Bayesian perspective in order to take advantage of reference distributions and also to quantify the uncertainty  for the measurements over a given period.  A Bayesian hierarchical model (BHM) also builds in more symmetry between the trend and anomalous signals. The trend is based on a multivariate normal distribution prior for the basis coefficients, and the anomaly uses a prior for the coefficients that is double exponential (Laplace). Besides providing a useful solution to our motivating application, the comparison between frequentist and Bayesian approaches is interesting in that both are able to accurately detect anomalous behavior and also have low false positive rates.  Both approaches are utilized in our application to offline monitoring at Urban Water Treatment Facility\footnote{This treatment facility has requested that its identity not be listed.} (UWTF). 

UWTF faces the challenge of providing a safe and ample supply of drinking water.  This is accomplished over a multistep process, which includes gravity flow through a filter consisting of anthracite, sand, and gravel. The performance of these filters is tracked with a time series variable known as {\it headloss}, which measures reduction in water pressure as the media becomes clogged. When headloss in a filter reaches a certain level, usually every one to two days, an operator shuts down the filter, and a backwash is initiated to clean the filter. 

The scale of the UWTF system precludes an operator from tracking in real time every filtration cycle on each of the multiple filters, and therefore, it is important to provide a statistical method to flag unusual filtration cycles for further examination in a timely, but retrospective manner.   Our analysis can be performed on the scale of seconds, while a backwash is performed over the course of hours, providing sufficient time to identify out of control operation before a filter resumes.  Because headloss is recorded on five minute intervals, it would be possible to achieve online monitoring in this application if desired.

\begin{figure}
    \centering
    \includegraphics[scale=.5]{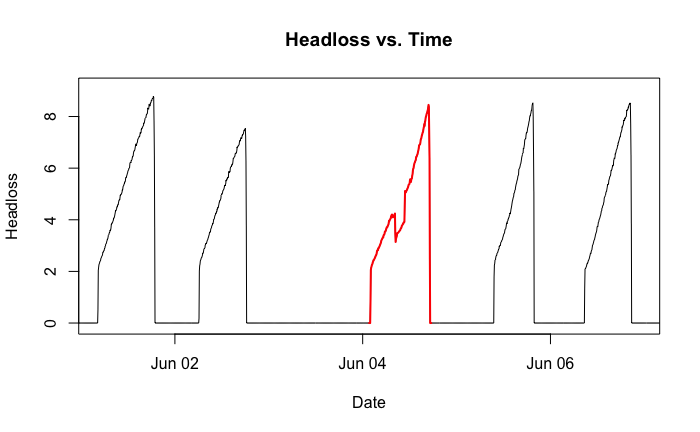}
    \caption{Five headloss cycles are seen here. The third is anomalous.}
    \label{headloss_over_time}
\end{figure}

\begin{figure}[H]
    \centering
    \includegraphics[scale=1]{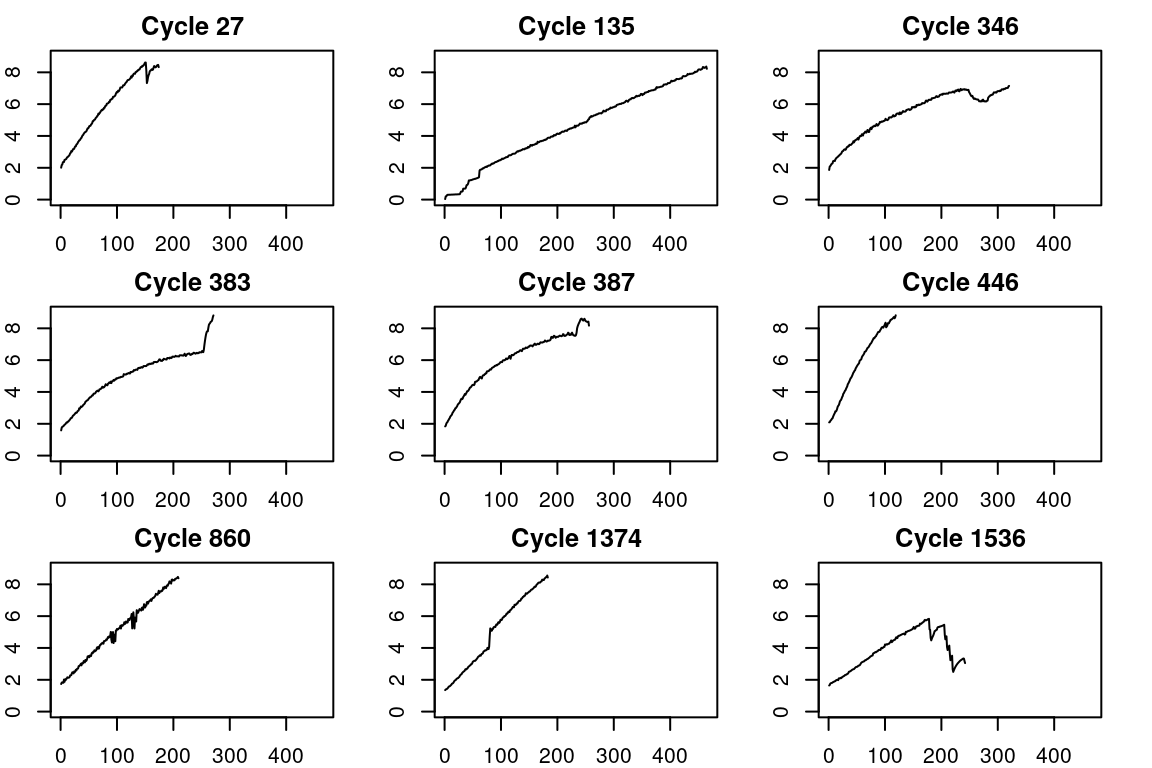}\\
    \includegraphics[scale=1]{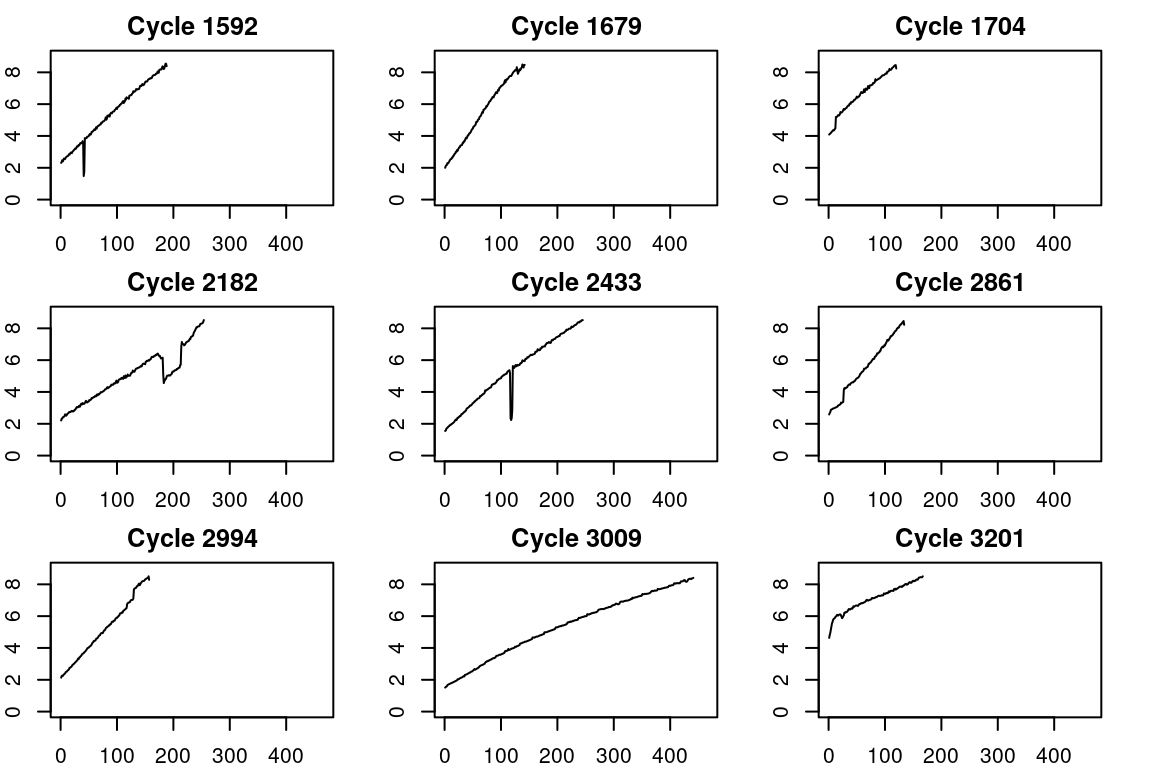}
    \caption{Additional anomalous cycles measuring headloss over time.  Each time unit represents one measurement taken every five minutes. See Figure \ref{zoomin} for a closer look at and explanation of cycle 3009. This cycle was not included in the key of anomalous cycles but was later discovered to have strange behavior.}
    \label{anomalous_examples}
\end{figure}

Headloss measurements in normal cycles increase steadily over time; however, cycle length, height, rate, and concavity vary between cycles making a number of methods of anomaly detection, such as profile monitoring, infeasible. The first two and the last two cycles in Figure \ref{headloss_over_time} are normal. The third cycle is anomalous.  Anomalous cycles can take a variety of forms but generally consist of at least one large jump in the measurements of headloss over time.  We refer to these jumps as change points, anomalies, or disturbances in the smooth, unknown trend.  A sample of anomalous cycles containing change points can be found in Figure \ref{anomalous_examples}.  

There are other ways in which a cycle can be outside of the norm, including being very short or not reaching the 8-8.5 units of headloss value before a backwash is initiated; however, those cycles can be easily found and flagged using other methods.  In contrast, our analysis looks for disturbances in the trend that can often be subtle.  Although the development of our approach is focused on this type of time series,  a general model that combines a GP with a non-Gaussian and sparse process may be useful in other applications. In particular, this method can be adapted to other change point detection problems in a time series. In fact, this type of separation is possible along with the ability to sample from the Bayesian posterior of either signal. 

In Section \ref{History}, we place this contribution within the literature for control and anomaly. In Section \ref{General Model}, we detail a general model, the relationship between cubic smoothing splines and GPs, and LASSO regression.  We then develop a model for curve fitting that combines the $L_1$ and $L_2$ penalties and outline an optimization algorithm for estimating the coefficients.  This frequentist approach requires choosing penalty values, and we adapt a machine learning idea for this task.  In Section \ref{bayes}, we adopt a Bayesian hierarchical approach and choose priors that are comparable to the $L_1$ and $L_2$ penalties in Section \ref{General Model}.  We then discuss adjustments that can be made which result in a full Gibbs sampler and overcome the confounding that often occurs in Bayesian models mixing fixed and random effects \citet{hodges2010adding}. In Section \ref{sim.study}, we test the robustness of the hybrid methods in a Monte Carlo study over a range of trends, jump sizes, anomaly locations, and noise conditions and then compare their effectiveness to two-step approaches of first detrending data and then using cumulative sum control charts (CUSUM), exponentially weighted moving average (EWMA), and a pruned exact linear time (PELTS) algorithm on the residuals.  In Section \ref{UWTF}, we compare the effectiveness of the two approaches in identifying anomalous cycles in the headloss data and present a general assessment of each approach.  Finally, in Section \ref{extensions}, we briefly discuss an extension to our model to test for other types of anomalous behavior.

\section{Anomaly Detection in Time Series}\label{History}

Methods of process control and anomaly detection date back to the introduction of Shewhart charts \citep{shewhart1931economic} and  CUSUM \citep{page1954continuous} used to detect changes in the mean of stationary time series. Many refinements to CUSUM have been made since then.  CUSUM can be adapted to detect multiple change points but requires a restart after the detection of each mean shift. Separately, EWMA \citep{hunter1986exponentially}, used widely in trading, is also used to detect subtle change in mean over time.  

Additional work to detect multiple change points in a time series include binary segmentation (BS), a method of cluster analysis \citep{scott1974cluster}, and segment neighborhood (SN), which searches for changes in model parameters between windowed segments of the data \citep{auger1989algorithms}.  Unlike BS, which is not guaranteed to provide the optimal solution, SN tests the entire solution space but has large computational cost.  The PELTS algorithm \citep{killick2012optimal} is an improvement on SN that provides an optimal solution for multiple change point detection in linear time.  

In addition to these segmentation approaches, there are kernel and Gaussian Process based methods such as \citet{lebarbier2005detecting} and \citet{saatcci2010gaussian}, the latter of which is an online approach. Bayesian approaches to multiple change point detection include \citet{barry1993bayesian},  \citet{green1995reversible}, and \citet{adams2007bayesian} and allow for probabilistic inference in multiple change points, a reversible jump in the dimensionality of the parameter space, and online monitoring, respectively.

All of the aforementioned approaches come with the assumption of stationarity and therefore when applied to non-stationary data require some form of detrending and/or transformation.  This two-step approach of first detrending the data and then applying a method is common throughout time series analysis.   Separation of these two steps is somewhat arbitrary and is an important critique.  In this paper, we seek to add to available methodologies by proposing an approach for multiple change point detection on non-stationary data that removes an unknown trend while detecting change points in a single step.

\section{ A Frequentist Approach to Identifying Anomalies}\label{General Model}
In order to identify the change points in any anomalous cycle, we consider our data to have come from the additive model
\begin{equation}\label{model}
y(t_i)=f(t_i)+g(t_i)+\epsilon_i
\end{equation}

\noindent where $y(t)$ is the observations, $f(t)$ is a smooth function, $g(t)$ is a rough function, $\epsilon_i \sim N(0,\sigma^2)$, and $\{t_i\}$ are $n$ time points.  The observation points are equally spaced in our application but are not required to be.  In our case, we identify the smooth function, $f(t)$, as the normal trend of the process over time, and the rough function, $g(t)$ represents possible anomalies that occur having abrupt change points.  In doing so we place only one assumption on the unknown trend, that it is smooth, and we define the change points as those locations in the time series where it is not.  We propose this additive model because in our water application and many others, the nature of the trend is relatively unknown.  The slope, concavity, and noise levels in headloss are affected by rainfall, season, temperature, and a number of operator choices and result in a number of odd features that are not always indicative of a problem.  Smoothness can extend beyond continuity and is discussed in Section \ref{extensions}. 
  
\begin{figure}[H]
    \centering
    \includegraphics[scale=.25]{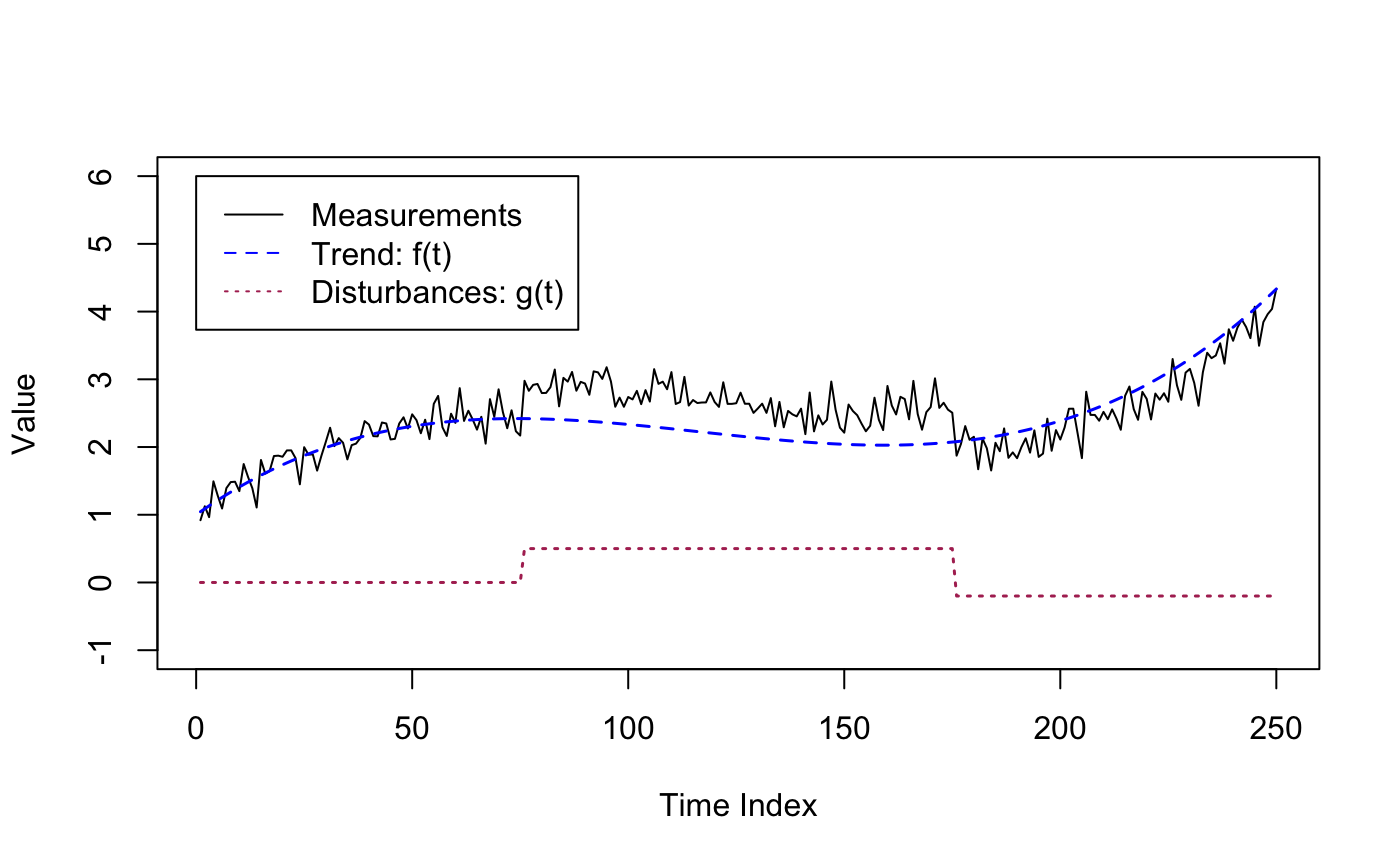}
    \caption{An example of the components in the additive model in Equation \ref{model}.}
    \label{Pedagogy}
\end{figure}

Figure \ref{Pedagogy} illustrates our model with a stylized anomaly pattern and hypothetical observations. The key to separating the smooth and rough components lies in our hybrid smoothing model for $f(t)$ and $g(t)$. We will represent $f(t)$ as cubic smoothing spline and $g(t)$  as being dominated by a sparse collection of step changes. Moreover, our BHM formulation has a posterior mode that is comparable to the hybrid smoother.

\subsection{\textbf{Cubic Smoothing Splines and  GPs}}\label{covariance}
Representing the cubic smoothing spline as a method of curve fitting via a smoothness penalty on the second derivative of the model was originally proposed by \citet{reinsch1967smoothing}.  The spline curve  is continuous in its first and second derivatives and is linear beyond its outer knots. We begin with the additive model omitting $g(t)$ as follows.
\[ y(t_i)=f(t_i)+\epsilon_i \]
 The cubic spline optimization can be written as
\begin{equation}
\begin{array}{rclcl}
\displaystyle \min_{f  \in {\cal H}} & \multicolumn{3}{l}{\sum_{i=1}^n(y(t_i)-f(t_i))^2 + \omega \int (f(t)'')^2dt}
\end{array}
\end{equation}
where $\omega$ is the smoothing parameter.

Formally,  ${\cal H}$ is a particular Hilbert space of functions but operationally can taken to be all functions where the smoothness penalty is finite.  
This solution can be characterized with a finite representation.
There is a set of  $n$ basis functions, $\{ \phi_k (t) \}$, and coefficients, $\{ c_k \}$, with 
$f(t) = \sum_{k=1}^n \phi_k (t) c_k $ and  the coefficients are found as 

\begin{equation}
\begin{array}{rclcl}
\displaystyle \min_{\bc} & \multicolumn{3}{l}{\| \by-\Phi \bc\|^2_2+ \omega \bc^T R \bc}
\end{array}
\end{equation}
\noindent
where $\Phi_{i,k} = \phi_k (t_i)$ and $R$ is a Gram matrix of the smoothing penalty applied to the basis functions 
\[ R_{j,k} =  \int  \phi_j(t)^{\prime \prime} \phi_k(t)^{\prime \prime} dt \]
 There are several common ways to choose these basis functions that will give equivalent results but may result in different degrees of conditioning in the matrix $\Phi$. In all cases, the spline is found as a form of ridge regression and the coefficients are found using the usual ridge regression formulation. Finally, we note that $\Phi$ has full rank, so it can be reparametrized in terms of the spline function values at the data locations as follows:
 
\begin{equation}
\begin{array}{rclcl}
\displaystyle \min_{\bbf } & \multicolumn{3}{l}{\| \by- \bbf \|^2_2+ \omega \bbf^T \Gamma \bbf  }
\end{array}
\end{equation}
\noindent
Here $\bbf_i = f( t_i)$, and  $\Gamma =  \Phi^{-T} R  \Phi^{-1} $. Some algebra shows that the minimizer in this form is
$ \hat{\bbf} = (I + \omega\Gamma)^{-1} \by$ and can be interpreted as a linear smoother applied to the time series in 
$\by$. A subtlety introduced in this version is that the penalty $\bbf^T \Gamma \bbf $ will be zero when $\bbf$ is a linear function, and so $\Gamma$ will  have a rank of $n-2$.

An important property of a cubic spline and many other splines is that they can be interpretted as a prediction from a Gaussian process model.  For clarity here, we focus on the stochastic interpretation of a cubic spline and prediction at the observation locations, although we note that these ideas hold in general for the family of multidimensional thin plate splines and prediction at arbitrary locations. 

Assume that $f(t) = P(t) + h(t)$, where $P(t)$ is a linear function in $t$, and $h(t)$ is a zero mean, Gaussian process with covariance function
\[ COV(h(t), h(t^*) ) =   k( t, t^*) \]
Define an $n \times 2$ regression matrix $\bX$ with the $i^{th}$ row given by  $[ 1, t_i]$. That is, $\bX$ is the usual regression matrix for fitting a linear trend.  In this setup, we have the usual linear GP model
\[ \by = \bbf +  \epsilon = \bX\bbeta + \bh +  \epsilon \] 
where $\bh_i = h(t_i)$. Now, let $K$ be the covariance matrix of $\bh$. It is a standard result that the best linear unbiased estimator for $\textbf{f}$ is given as 
\[ \hat{\bbf} =  \bX\hat{\bbeta} +  K( K + \sigma^2 I)^{-1} ( \by - X \hat{\bbeta} ) \]
where $\hat{\bbeta}$ is the generalized least squares estimator of $\bbeta$.  Under this model, the error vector, $\bbf - \hat{\bbf}$, will be  distributed $MN\sim(\mathbf{0},\Omega)$ where $\Omega$ depends on $\sigma^2$, $X$,  and $K$.   The surprising result is that one can also show that the estimator, $\hat{\bbf}$, is also obtained as the solution to the minimization problem:
\begin{equation}
\begin{array}{rclcl}
\displaystyle \min_{\bbeta, \bh } & \multicolumn{3}{l}{\| \by- \bX \bbeta - \bh \|^2_2+ \omega \bh^T K^{-1} \bh  }
\end{array}
\end{equation}

\noindent Using a specific choice of covariance function, $k(\cdot,\cdot)$,  this  inference for a Gaussian process model will be the \textit{same} as a cubic smoothing spline and was an early landmark result connecting numerical splines to statistics (\citet{wahba1990spline}, p.39).   It can be implemented with the {\tt Tps.cov()} function in the R package \textbf{fields} (\citet{fields}). Thus, we have connected the initial smoothing problem for functions to an estimator that assumes a stochastic process model for the unknown function.  This equivalence is useful when interpreting the BHM in Section 3.

The last elaboration needed for our work is the  extension of this correspondence to a complete Bayesian model where all unknown quantities have stochastic, prior distributions.  Suppose one places a prior on $\bbeta$ as $MN(\mathbf{0}, bI)$ and conditions  on all other statistical parameters, e.g. $\sigma^2$ and $K$. The conditional distribution of $\textbf{f}$ given $\by$ is, again, multivariate normal. Moreover in the limit as $b \rightarrow \infty$, this conditional distribution will converge to $MN( \hat{\bbf}, \Omega) $ with the  same mean and covariance as detailed above.

\subsection{LASSO  and a Non-Gaussian Model.}
\citet{tibshirani1996regression} was the first to introduce the LASSO as a method of constraining the estimates of the linear regression coefficients via an $L_1$ penalty as
\begin{equation}
\begin{array}{rclcl}
\displaystyle \min_{\bgamma} & \multicolumn{3}{l}{\| \by-\bQ\bgamma \|^2+ \lambda\|\bgamma\|_1}
\end{array}
\end{equation}
where $\lambda$ is the regularization parameter.

Here, $\bQ$ denotes the usual form for covariates in a linear model where $E(\by) = \bQ\bgamma$. 

\noindent 
A strength of the LASSO model as a frequentist method is that the parameter estimates can be identically zero where the number of zeroes is controlled by the size of $\lambda$. In our application, however, covariate selection is not as important as the effect of down weighting many of the components of $\bgamma$ and leaving a sparse number with larger values. 

In parallel with the cubic spline development, let $\boldsymbol{\psi}_k(t)$ be a basis, and expand $g$ as 
\[ g(t) = \sum_{k=1}^n \boldsymbol{\psi}_k(t) \bgamma_k \]
To model steps in $g$,  we take the basis to be 
\begin{equation}
\boldsymbol{\psi}_k(t) = 
\left\{ \begin{array}{ll}
0 & \mbox{ for } t < t_k \\
1 & \mbox{ for  } t  \ge t_k \\
\end{array}
\right.
\end{equation}
Let $\Psi_{i,k} = \boldsymbol{\psi}_k(t_i)$ and substitute this matrix for $\bQ$ in the generic LASSO formulation above. 
From a frequentist perspective, we use the $L_1$ penalty to seek a model with a limited number of steps in $g$. Although this simple device works well for the water filtering application, one could also extend this to more complicated anomalies by modifying the basis functions.  More about this flexibility will be discussed in Section 5. 

\citet{tibshirani1996regression} also suggested that the LASSO could be interpreted as a double-exponential (Laplace) prior on $\bbeta$ in the Bayesian framework.  \citet{figueiredo2003adaptive} expanded on this work and was one of the first to publish a Bayesian hierarchical model using Laplace priors. \citet{park2008bayesian} followed, proposing a Bayesian LASSO with a full Gibbs sampler.

\subsection{Combined $L_2$ and $L_1$ Hybrid Curve Estimator}\label{frequentist}

Based on the preceding development, we consider the combined model as 
\[ \by=\Phi \bc + \Psi \bgamma + \epsilon \]
and the combined optimization for these coefficients as 
\begin{equation}\label{eq4}
\begin{array}{rclcl}
\displaystyle \min_{\bgamma, \bc} & \multicolumn{3}{l}{\|y-(\Phi \bc + \Psi \bgamma)\|^2+ \lambda\|\bgamma\|_1 + \omega \bc^TR\bc}
\end{array}
\end{equation}
where $R$ is the cubic smoothing spline penalty matrix discussed in Section \ref{covariance}.  Previous work by \citet{shen2014detection} proposed a similar use of LASSO, modeling the trend with a linear function. Here, however, we allow the unknown trend to be modeled by any smooth function. This new expanded model is not a particularly easy optimization problem to solve as stated. However, one can take advantage of the fact that minimization over $\bc$ in the quadratic terms has an analytic solution. Substituting this result back into (\ref{eq4}) reduces  the program  to a straight $L_1$ optimization over only $\bgamma$.  Accordingly, the minimization with respect to $\bc$ is 
\[ \hat{\bc}=(\Phi^T\Phi+\omega R)^{-1}\Phi^T(\by- \Psi \bgamma) \]

\noindent Setting $S(\omega) =(\Phi^T\Phi+\omega R)^{-1}$ and $\bu= (\by-\Psi \bgamma)$ 
gives  $\hat{\bc}=S\Phi^T\bu$, and we have the reduced problem in $\bgamma$ as  
\begin{equation}
\begin{array}{rclcl}
\displaystyle \min_{\bgamma} & \multicolumn{3}{l}{\|\bu-\Phi S(\omega) \Phi^T\bu\|^2_2+ \lambda\|\bgamma\|_1+\omega (\bu)^T\Phi S(\omega) ^TRS(\omega)\Phi^T\bu  }
\end{array}
\end{equation}

\noindent Some straightforward algebra gives a surprisingly simple reduction of
\[ \min_{\bgamma}  {(\bu)^T(I-\Phi S(\omega)\Phi^T )\bu+\lambda\|\bgamma\|_1} \]

\noindent Letting $\bW(\omega)$ be the symmetric square root matrix of  $I-\Phi S(\omega) \Phi^T$, we arrive at the following conventional LASSO problem with optimization details found in Appendix \ref{fista}: 
\begin{equation}
\label{reducedLASSO}
 \min_{\bgamma} {\| \bW(\omega) \by -  \bW(\omega)  \Psi \bgamma \|_2^2+\lambda\|\bgamma\|_1} 
\end{equation}

Given the solution to (\ref{reducedLASSO}), one then uses the relationship with $\hat{\bc}$ given above to find the spline component.  An important feature in this derivation is the presence of the penalty parameters $\omega$ and
 $\lambda$ throughout, and for this reason, we have carried the dependence on the spline parameter, $\omega$,  forward to this last expression. The main statistical challenge is to estimate these penalty parameters based on the data, and the computational challenge is to recompute the solution to the optimization problem for each  new combination of  $(\omega, \lambda)$.  These aspects are addressed in the following sections.

\subsection{Choosing Values of $\lambda$ and $\omega$} 

It is necessary to choose values of $\lambda$ and $\omega$ that balance the signals and noise and also distinguish between $f$ and $g$.  The goal is to determine values that simultaneously result in a small RMSE, a sparse $\bgamma$, and a smooth trend.   In the case of a smoothing spline or GP, this is typically done by cross validation (CV)  or maximum likelihood, and in the case of the LASSO penalty, one uses an information criterion or CV.  The combination of the LASSO and smoothing spline also suggests using CV.  However, we have reservations with this approach.  In cross validation, an estimate for the removed point is made at the omitted locations.  For smooth functions, this is straightforward, but for an anomaly taking the form of a step function, a removed point at or adjacent to the jump is problematic. For this reason, we considered two methods in choosing $\lambda$ and $\omega$ that are distinct from CV.  The first is a modified machine learning approach, and the second is a version of the Akaike's Information Criterion (AIC) (\citet{bozdogan1987model}).  

\subsubsection{The Elbow - A Machine Learning Algorithm}
 
We modify an approach based on a machine learning method by \citet{5961514} that provides a good choice of $\lambda$ and $\omega$ with the aim of simultaneously achieving a small root mean squared error (RMSE), a parsimonious signal, and a small number of non-zero $\bgamma$.
For any specific $\lambda$ and $\omega$,
define the RMSE as
\[ E(\lambda, \omega) = \sqrt{ (1/n) \sum_{i=1}^n ( y_i - \hat{f}(t_i) - \hat{g}(t_i) )^2 } \]
 of  the fit to the data using the hybrid model, and also let $E_0$ denote the RMSE for the base model with just a linear trend and fit by least squares. 
Let $H_2(\lambda)$ be the number of nonzero elements in $\hat{\bgamma}$, and then $N(\lambda, \omega)$ denotes the effective degrees of freedom for a given pair of penalty parameters defined as 
$$N(\lambda,\omega)=H_1(\omega)+H_2(\lambda)$$
with $H_1(\omega)=tr(\Phi(\Phi^T\Phi+\omega R)^{-1}\Phi)$.  For justification, recall
\begin{align}\label{hats} 
	\hat{\by}&=\Phi\hat{\bc}+\Psi\hat{\bgamma} \nonumber \\
	&=\Phi (\Phi^T\Phi+\omega R)^{-1}\Phi^T(\by- \Psi \hat{\bgamma}) +\Psi\hat{\bgamma} \nonumber \\
	&=\Phi (\Phi^T\Phi+\omega R)^{-1}\Phi^T\by + (I-\Phi (\Phi^T\Phi+\omega R)^{-1}\Phi^T)\Psi\hat{\bgamma} \\
	&=H_1\by + H_2\hat{\bgamma} \nonumber
\end{align}  
where $H_2(\lambda)=(I-\Phi (\Phi^T\Phi+\omega R)^{-1}\Phi^T)\Psi$.

The effective degrees of the freedom in a model with a smoothing penalty is the rank of the hat matrix, which can be estimated with the sum of the its diagonal (\citet{tibshirani1987local}).  In this model, however, we have the hat matrix, $H_1$, and a second matrix, $H_2$, we need to consider.  It is clear from Equation \ref{hats} that $H_1$ and $H_2$ are orthogonal to one another, and therefore it is reasonable to estimate the effective degrees in the model as the sum of effective degrees of freedom in $H_1$ and the degrees of freedom associated with $H_2$.  \citet{zou2007degrees} proposed using the number of active coefficients in a LASSO model as an unbiased estimate of the degrees of freedom in $H_2$.

Because $N(\lambda,\omega)$ represents the degrees of freedom in the model, we omit any choices of $\lambda$ and $\omega$ that result in $N(\lambda,\omega)>n$.
Now, consider the three dimensional coordinates 
\[ \bv(\lambda, \omega) = ( N(\lambda, \omega)/ n ,  S(\lambda)/n, E(\lambda, \omega)/ E_0 ). \]

\noindent By this normalization, $0 \le N/n \le 1$ and $0 \le S/n \le 1$.  The base spline model also includes a linear trend. Thus,  the  RMSE will be decreasing as a function of $N$ and $S$, and $E(\lambda, \omega)/ E_0$ will also be bounded between 0 and 1.   Taken together,  $\bv$ will be under the standard $2-simplex$ in the positive octant of $\mathcal{R}^3$. 
Estimates for $\lambda$ and $\omega$ are now determined by searching $\bv(\lambda, \omega)$ over a grid of values and finding the  point that is farthest from the simplex plane. The intuition behind this choice is that it makes the RMSE small while keeping both the overall model complexity ($N$) and the sparsity of $\bgamma$ ($S$) small. The distance to the simplex plane for a given $(\lambda,\omega)$ is
$$\text{distance}(\lambda,\omega) \propto 1 - \sum_{k=1}^3 \bv(\lambda,\omega)_k$$

Typically, the ``elbow" point in a function for RMSE  is identified as a good choice for a penalty parameter and  is found quantitatively as the maximum in the  second derivative of the RMSE curve. However, the sparsity of $\bgamma$  results in  $S(\lambda)$ not being twice differentiable, so instead the elbow point is inferred by the farthest distance from the simplex plane. 

\begin{figure}[H]
    \centering
    \includegraphics[scale=.3]{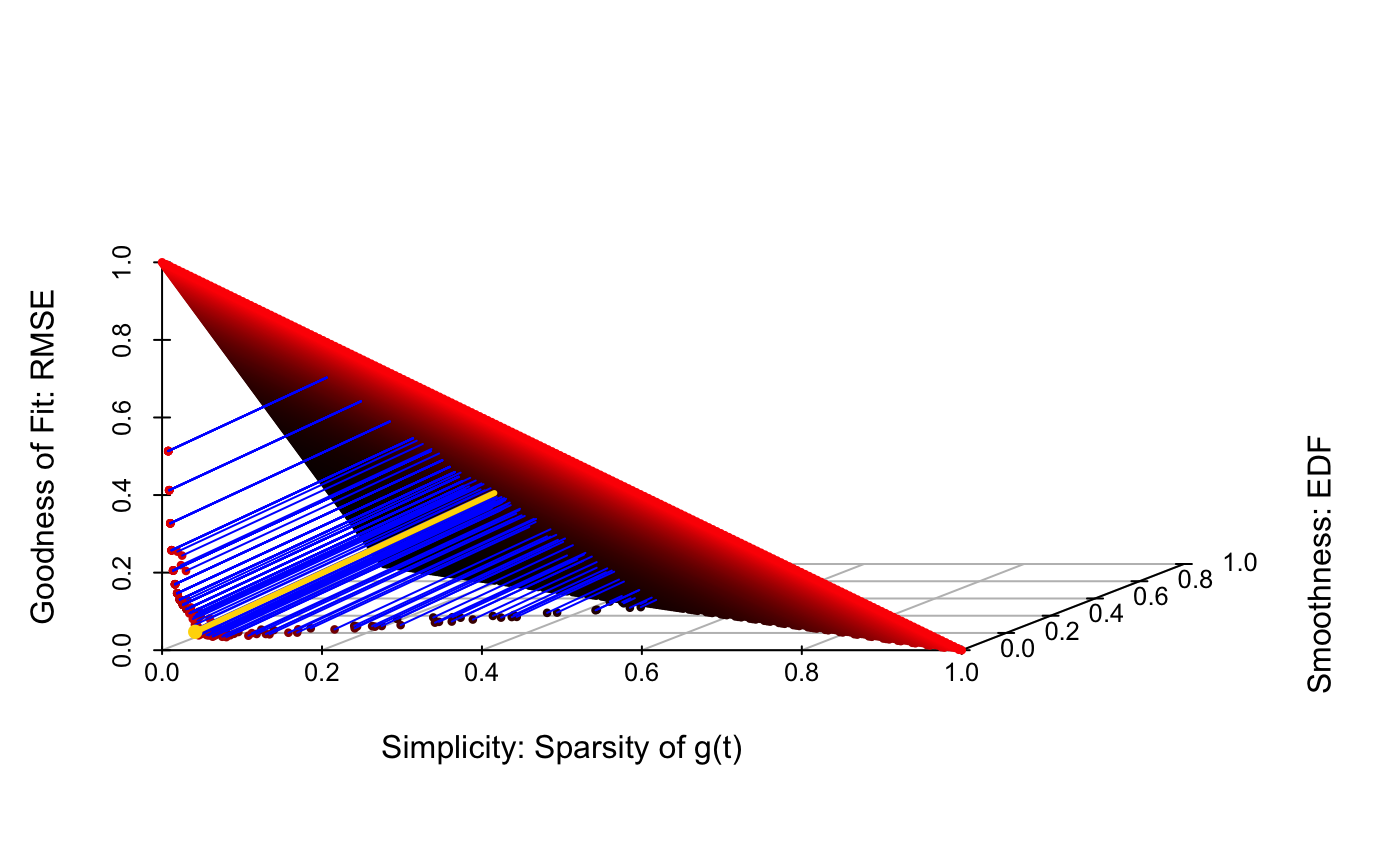}
    \caption{The lines in blue represent the perpendicular distances to the 2-simplex from the surface $V(\lambda,\omega)$.  The furthest point (gold) is taken to be the estimates of $\lambda$ and $\omega$.}
    \label{lambda_and_omega}
\end{figure}

\subsubsection{Akaike Information Criterion} 

It is common among change point methods to search for multiple change points by adding a penalty to the cost function \citep{killick2012optimal}.  This penalty is often linear in the number of change points, and examples include the Akaike Information Criterion (AIC) and the Bayesian Information Criterion (BIC).  Here, we adapt a version of  the corrected Akaike Information Criterion (AICc), that while not linear, imposes an upper limit on the number of change points. We give each combination of $\lambda$ and $\omega$ a score using
$$AICc (p)= log(\hat{\sigma}^2_p)+\frac{n+p}{n-p}$$
where $\hat{\sigma}^2_k=\frac{SSE(k)}{n}$, and $p=N(\lambda,\omega)$.  We implement this criterion using the same estimate for effective degrees of freedom used with the elbow approach.

\subsection{Identifying Anomalous Cycles}
\begin{figure}[H]
    \centering
    \includegraphics[scale=.4]{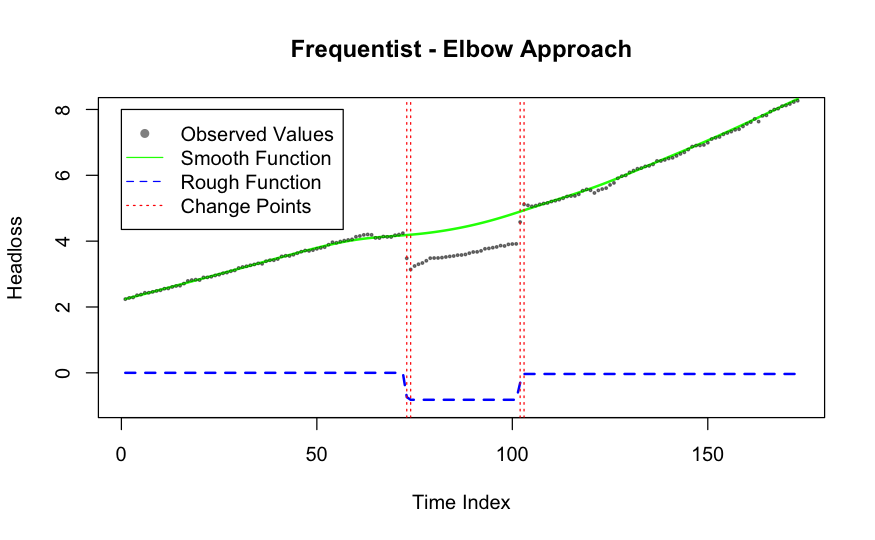}
    \caption{The decomposition of the third cycle in Figure \ref{headloss_over_time}.  The model found the time indices $73,74,102$ and $103$ to be change points based on the elbow method for determining $\lambda$ and $\omega$.}
    \label{full decon}
\end{figure}

Figure \ref{full decon} shows the results of estimating $\lambda$ and $\omega$ and the corresponding signals for the third cycle in Figure \ref{headloss_over_time} using the elbow method.  The ordinary path of the cycle is well modeled by the cubic smoothing spline, and the anomalous portion of the cycle is seen in the rough function, $g(t)$.  Change points are found for this frequentist approach as those $\bgamma$ with values that are non-zero.  

These frequentist (hybrid) approaches to identifying change points have proven very effective. However, due to the difficulty of quantifying uncertainty, we develop an alternative, fully Bayesian approach where $\lambda$ and $\omega$ can be identified as parameters in a hierarchical model.  This switch to a Bayesian framework will also provide error estimates on the model coefficients and thus quantify the uncertainty in the detection process.

\section{A Bayesian Approach to Identifying Anomalies} \label{bayes}

We now propose a Bayesian hierchical model (BHM) for identifying disturbances where the penalties in the hybrid approach are interpreted as the log of prior densities while the mode of the log posterior distribution is related to the solution to the optimization problem in Section \ref{frequentist}. In this way, the two can be compared.

\subsection{A Hierarchical Model}
We propose the top level to be the same observational model, but now, assume a Laplace prior for $\bgamma$ and a multivariate normal prior on $\bc$, as follows:
\begin{align}
y | \bc,\bgamma,\sigma^2 &\sim N_n(\Phi \bc + \Psi\bgamma, \sigma^2 I_n) \nonumber \\
[\bgamma|\sigma^2,\lambda ] &= \left(\frac{\lambda}{2\sqrt{\sigma^2}}\right)^{n-1} e^{-\frac{\lambda}{\sqrt{\sigma^2}}||\bgamma||_1} =\prod_{j=1}^{n-1} \frac{\lambda}{2\sqrt{\sigma^2}}e^{-\frac{\lambda} {\sqrt{\sigma^2}}|\bgamma_j|}\nonumber \\
[\bc | \omega]  & \propto \omega^{1/2}exp(-\frac{\omega}{2}\bc^TR\bc) \\
[\sigma^2] &\propto 1 \nonumber\\
\lambda &\sim gamma(\alpha_{\lambda},\beta_
{\lambda}) \nonumber \\
\omega &\sim gamma(\alpha_{\omega},\beta_
{\omega}) \nonumber  
\end{align}

\noindent Notice that the prior on $\bgamma$ is an uncorrelated multivariate $Laplace(0,\frac{\sqrt{\sigma^2}}{\lambda})$.  Additionally, the prior on $\bc$ is a multivariate Gaussian distribution; however, $R$ is not full rank. The rank deficiency of $R$ poses a problem that needs to be considered.  One way to address this problem is to utilize the elegant GP representation of the smoothing cubic spline discussed in Section \ref{covariance}.  The smooth trend, $\Phi \bc$, becomes $X\bbeta + \bg$ with priors $[\bbeta] \propto 1$ and $\bg \sim N(0,K/\omega)$.

These priors work to achieve sparseness in $\bgamma$ and smoothness for the trend. Finally, we use an improper prior on $\bbeta$ to be consistent with the BLUE frequentist estimator. Following the terminology in spatial statistics we refer to $
\bbeta$ as a {\it fixed effect} even though it is a random quantity in the BHM.

\subsubsection{Reformulating the $\bgamma$ Parameter} 
Unfortunately, the prior on $\bgamma$ is not conjugate, and if this model is implemented in its current form, sampling $\bgamma$ would require a Metropolis-Hastings step. Our goal is to formulate this model so that all sampling involves Gibbs updates. 
By adding an additional level  to this hierarchy with some extra parameters, it is possible to avoid a Metropolis-Hastings update on $\bgamma$, thereby achieving a Gibbs sampler but functionally maintaining the Laplace prior.  Here, we use the insightful result of  \citet{andrews1974scale} and utilized by \citet{park2008bayesian} where a Laplace distribution can be written as an exponential scale mixture of Gaussian distributions, as follows: 
\begin{equation}\label{laplace}
\frac{a}{2}e^{-a|z|}=\int_0^\infty \frac{1}{\sqrt{2\pi s}} e^{-z^2/(2s)} \frac{a^2}{2} e^{-a^2s/2}ds, \; \; \; a>0
\end{equation}

\noindent Using this representation and the GP version of the cubic smoothing spline results in the equivalent hierarchical model 
\begin{align}
y | \bbeta,\bg,\bgamma,\sigma^2 &\sim N_n(X\bbeta + \Psi\bgamma+ \bg, \sigma^2 I_n) \nonumber \\
\bgamma|\sigma^2,\tau_1^2,...,\tau_{n-1}^2 &\sim N_{n-1}(0_{n-1},\sigma^2F) \nonumber 
\text{  where } F = diag(\tau_1^2,...,\tau_{n-1}^2) \\
\tau_i^2 |\lambda^2 &\sim exp(\frac{\lambda^2}{2} ) \text{  for $i=1,...,n-1$} \nonumber \\
\bg |\omega &\sim N_n(0_n, \frac{1}{\omega} K) \nonumber \\
[\sigma^2] &\propto 1 \nonumber\\
\lambda^2 &\sim gamma(\alpha_{\lambda^2},\beta_
{\lambda^2}) \nonumber \\
\omega &\sim gamma(\alpha_{\omega},\beta_
{\omega}) \nonumber \\
[{\bbeta}] &\propto 1 \nonumber \\  
\nonumber
\end{align}

\noindent Here, the second and third levels give the same conditional distribution for $\bgamma$ as the second line of the first hierarchy, and a proof of this fact is given in Appendix \ref{LaplaceIdentity}.

\subsubsection{Reducing Confounding of Fixed and Random effects}

The BHM in $3.1.1$ suffers from an additional problem.  It does not mix well due to confounding between the fixed and random effects (\citet{hodges2010adding}). This problem is a well known feature in these types of BHMs and is distinct from the frequentist/optimization where modfications are not needed.  The introduction of the Gaussian process into this model is known to result in spatial confounding with the fixed effects in $X$.  Not accounting for this confounding results in poor mixing of $\bbeta$, $\bg$, and $\bgamma$ and a general failure for the MCMC chains to  converge.  The process of orthogonalizing the fixed effects with the random effects and then (nearly) orthogonalizing the Gaussian process, $\bg$, with the discontinuous stochastic component, $\Psi \bgamma$, 
is accomplished with projection matrices based on $\bX$ and $\Psi$.  In the process, the key elements $\bbeta$, $\bg$, and $\bgamma$  are reparametrized in a way that improves mixing and convergence. A novel element in this implementation is the use of a regularized version of the projection for the anomaly basis function to make the computations numerically stable.  

Let $\bP_X=X(X^TX)^{-1}X^T$ . Then,
\begin{align}
    \by&=X\bbeta+\Psi \bgamma+\bg + \epsilon \nonumber \\
    &=X\bbeta+P_\bX(\Psi \bgamma+\bg) +(I-P_\bX)(\Psi \bgamma+ \bg)+ \epsilon \nonumber\\
     &= X\bbeta^*+ \Psi^*\bgamma + (I-P_\bX)\bg + \epsilon \nonumber
\end{align}
with  the reparameterization $ \bbeta^* =\bbeta + (X^TX)^{-1}X^T(\Psi \bgamma+ \bg)  $
and $\Psi^*=(I-P_X) \Psi$, which is a projection of the anomaly basis functions onto the space orthogonal to the linear trend.

We now use the same operation to further improve sampling, letting 
 \[ P_{\Psi^*}=\Psi^*({\Psi^*}^T\Psi^*+I\delta)^{-1}{\Psi^*}^T\]
  where $\delta I$ functions as a small regularization term, on the order of single precision zero  (e.g., $\delta=10^{-8}$) because $\Psi^*$ is computationally singular. Thus, we have
\begin{align}
   \by &= X\bbeta^*+ \Psi^*\bgamma + (I-P_X)\bg + \epsilon \nonumber\\
     &= X\bbeta^*+ \left( \Psi^*\bgamma + P_{\Psi^*}(I-P_\bX)\bg \right) + (I-P_{\Psi^*})(I-P_\bX)\bg \nonumber\\
    &= X\bbeta^*+ \Psi^* \bgamma^* + H \bg
\end{align}
with the final reparametrizations 

 $\bgamma^*=\bgamma + J\bg$,  $J=({\Psi^*}^T\Psi^*+I\delta)^{-1} {\Psi^*}^T (I-P_\bX)$, and $H=(I-P_{\Psi^*})(I-P_X)$

Summarizing these modifications we arrive at an ``orthogonalized"  BHM  (OBHM) that is equivalent to the first model proposed, up to the approximation for $P_{\Psi^*}$,  but it is much more efficient to apply Gibbs sampling and a Markov Chain Monte Carlo (MCMC) algorithm. It is
\begin{align}
y|\bbeta^*,\bg,\bgamma^*,\sigma^2 &\sim N_n(X\bbeta^* + \Psi^*\bgamma^*+ H \bg, \sigma^2 I_n) \nonumber \\
\bgamma^*|\sigma^2,\tau_1^2,...,\tau_{n-1}^2,\bg &\sim N_{n-1}(J\bg,\sigma^2F) \nonumber 
\text{  where } F = diag(\tau_1^2,...,\tau_{n-1}^2) \\
\tau_i^2 |\lambda^2 &\sim exp(\frac{\lambda^2}{2}) \text{  for $i=1,...,n-1$} \nonumber \\
\bg |\omega &\sim N(0,\frac{1}{\omega} K) \nonumber\\
\omega &\sim \text{gamma}(\alpha_{\omega},\beta_{\omega}) \nonumber\\
\lambda^2 &\sim \text{gamma}(\alpha_{\lambda^2},\beta_
{\lambda^2}) \nonumber \\
[\bbeta^*] &\propto 1 \nonumber \\
[\sigma^2] &\propto 1 \nonumber
\end{align}

\noindent The full posterior conditional distributions for the model are found in Appendix \ref{posteriors}.

\subsection{\textbf{Convergence, Priors, and Initial Values}}

We implemented a uniform prior on $\bbeta$ so that the combination of the linear trend identified by $X\bbeta$ and the GP, $\bg$, is precisely a cubic smoothing spline. This equality was discussed in Section \ref{covariance}.  A uniform prior on $\sigma^2$ was chosen for the sake of simplicity.  We noticed, however, that a loosely informative inverse gamma prior on $\sigma^2$ results in a higher effective sampling rate.  We decided on loosely informative priors on  $\lambda^2$ and $\omega$, making use of the analysis from the hybrid approach.  These loosely informative priors are relatively flat, but they place greater likelihood near the mean of each parameter gathered from the frequentist approach.  Improper, uniform priors were tested for $\lambda^2$ and $\omega$ and rejected because they greatly reduced the effective sample sizes in each chain to the point that the MCMC struggled to converge.   

The initial values of $\bgamma$ were chosen by finding $z_i=|y_{i+1}-y_i|$ for $i=2,..,n$ and the $95^{th}$ percentile of these values, $z_{0.95}$.  We then defined $\bgamma$ as  

\begin{equation}
\bgamma_i = 
\left\{ \begin{array}{ll}
0 & \mbox{ for } z_i < z_{0.95} \\
z_i & \mbox{ for  } z_i  \ge z_{0.95}\\
\end{array}
\right.
\end{equation}

\noindent Given the initial guess of the disturbances, $\bgamma$, natural starting values for $\bbeta$ are the OLS coefficient estimates regressing $y-\Psi \bgamma$ on $X$.  We finally set the initial values of $\lambda^2$, $\omega$, and all $\tau^2_i$ to the mean values of their distributions and sampled $\sigma^2$ first in our MCMC. Convergence and mixing happen quickly in this formulation, requiring a burn-in of less than $200$ iterations and chains of only $1000$ dependent samples.

\subsection{\textbf{Identifying Disturbance Locations}}
\begin{figure}[H]
    \centering
    \begin{minipage}{.5\textwidth}
      \centering
      \includegraphics[scale=.27]{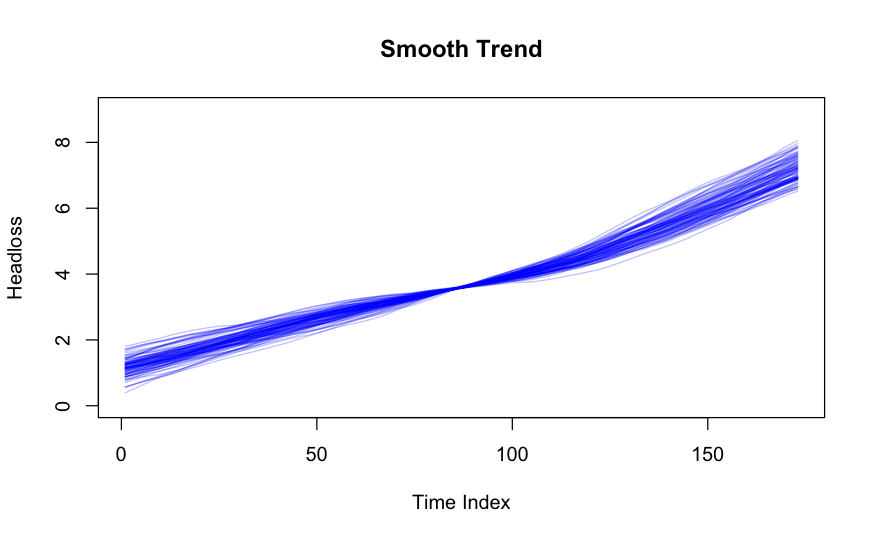}
    \end{minipage}%
    \begin{minipage}{.5\textwidth}
      \centering
      \includegraphics[scale=.27]{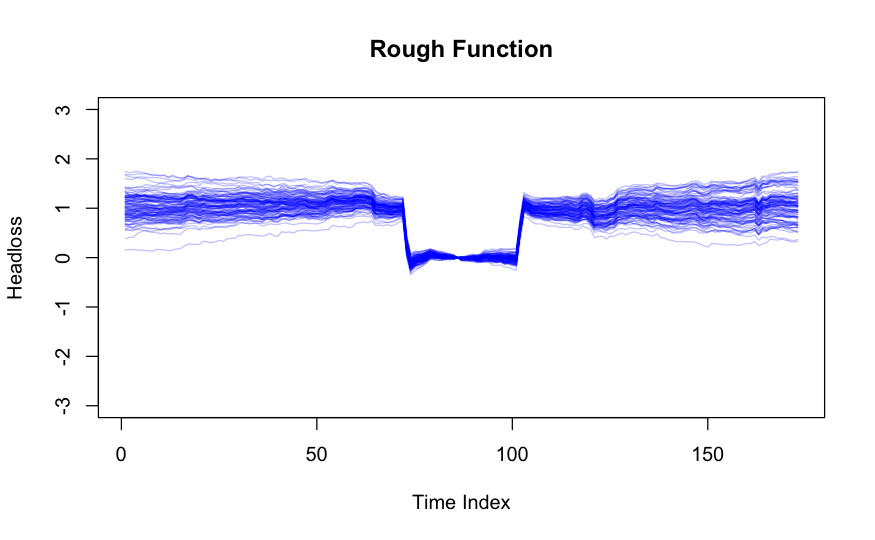}
    \end{minipage}
    \caption{The two plots are the decomposition of the third cycle in Figure \ref{headloss_over_time} using the Bayesian GP approach.  The left plot contains posterior predictions of $X\bbeta+\bg,$ and the right contains posterior predictions of $\Phi \bgamma$.}
    \label{fulldecon2}
\end{figure}

Figure \ref{fulldecon2} shows $50$ posterior predictions from the MCMC modelling the third cycle in Figure~\ref{headloss_over_time} with the smooth portion of the model on the left and the rough portion on the right. Again, we see a clear separation between the two.  An alternative form of $\Psi$ discussed in Appendix \ref{ConstructPsi} was used to reduce the size of the pointwise credible intervals of the $f(t)$ and $g(t)$ at the edges. This change has no effect, however, on the posterior credible intervals of $\bgamma$.

\begin{figure}[H]
    \centering
    \includegraphics[scale=.28]{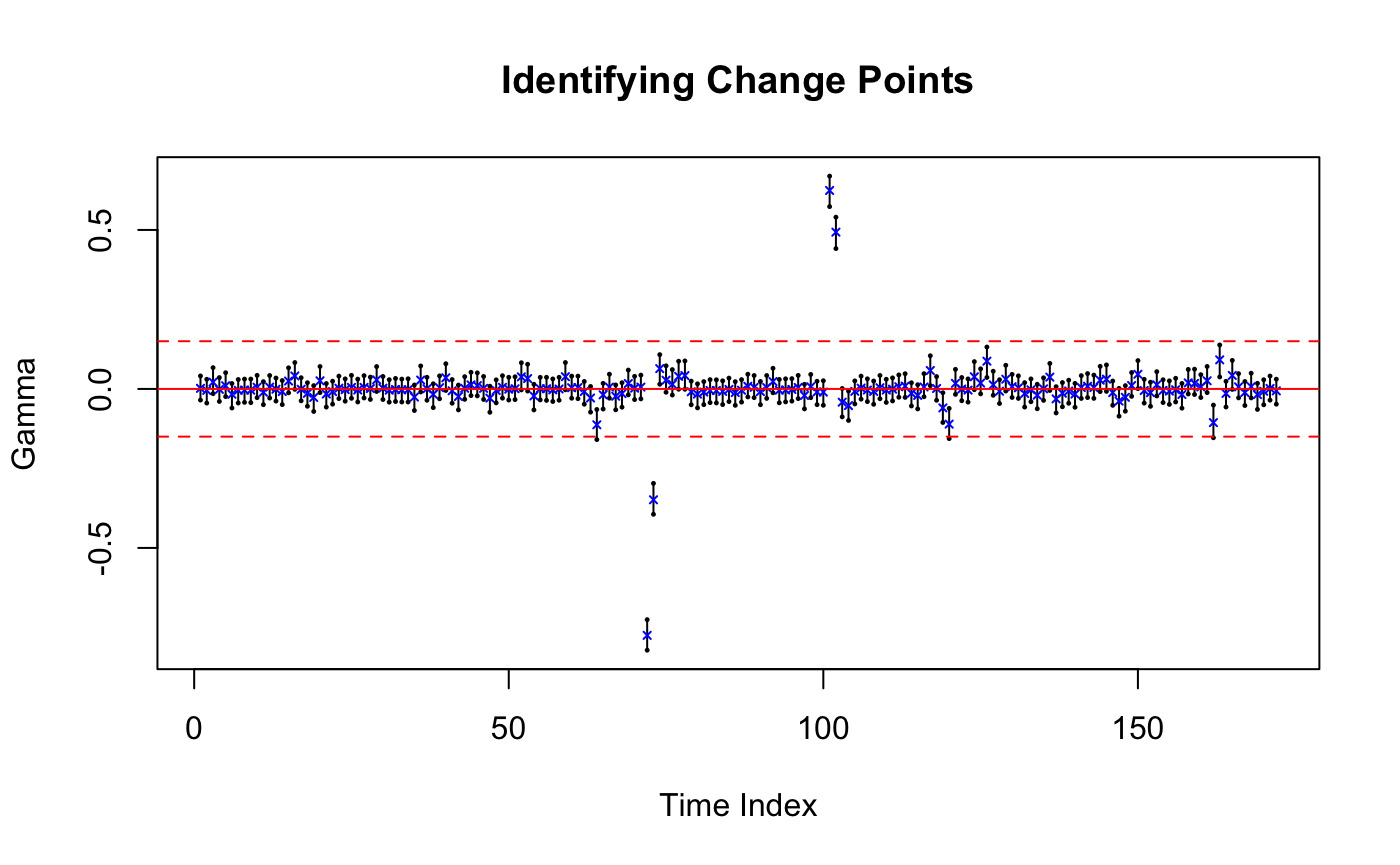}
    \caption{Displayed are the $95\%$ credible intervals of each $\bgamma$ coefficient.  The dotted lines are placed at $\pm 0.15$.  Any locations at which the credible interval does not include zero and the mean is outside of the dotted lines is flagged as a relevant change point.  This approach, similar to the hybrid approach, found the time indices $73,74,102$ and $103$ to be change points.}
    \label{gamma2}
\end{figure}

Figure \ref{gamma2} shows the $95\%$ credible intervals of each $\bgamma$ coefficient.  These posterior results are then used in combination with an application specific threshold to identify disturbances while minimizing false positive results.  UWTF initially choose a threshold of $0.15$ units headloss because smaller anomalies were not discernible by visual inspection.

\section{Simulation Study}\label{sim.study}
\begin{figure}[H]
    \centering
    \includegraphics[scale=.25]{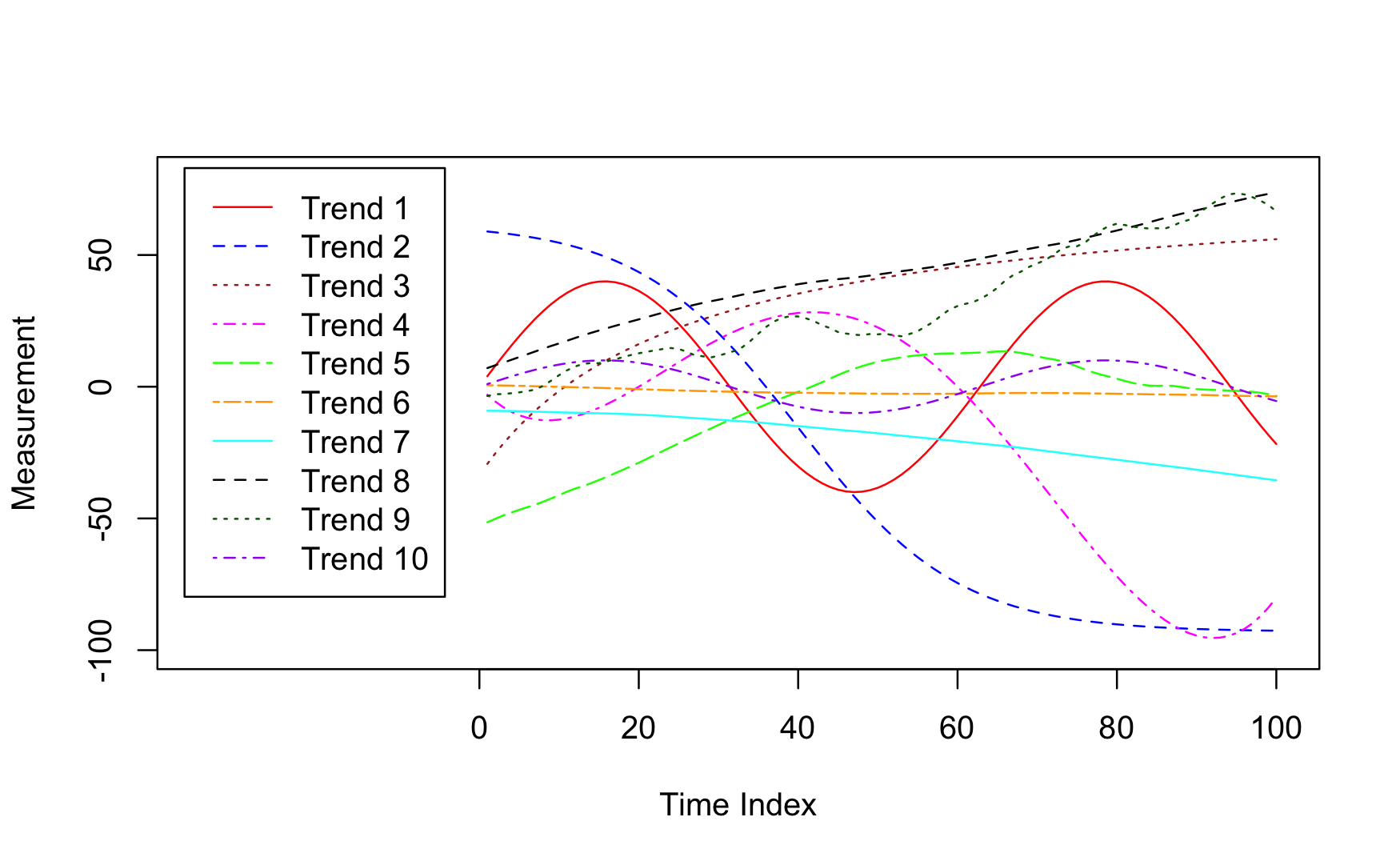}
    \caption{Ten trends of various complexity over which the simulation study is performed.}
    \label{trends}
\end{figure}

The hybrid approach to multiple change point detection was developed for use in our water filter application. However, wishing to demonstrate its effectiveness across a range of applications, we ran a simulation study over a range of data and conditions.  We tested the one step hybrid AICc and elbow approaches over ten trends (Figure  \ref{trends}) and multiple levels of noise, jump size, and change point frequency (0, 1, or 2). Each set of condition ran 200 times, and the location of the jumps were sampled randomly over the support.  Seeking to understand how trend features affect change point detection, the derivative and second derivative were recorded at the location of each jump.  Finally, to demonstrate the importance of this approach, we simultaneously test two-step CUSUM, EWMA, and PELTS techniques, first fitting a cubic smoothing spline to the data and second searching the residuals for change points.  CUSUM was set with a decision interval of $4\sigma$ and slack of $0.5\sigma$.  It was reset after each violation occured to allow for detection of multiple change points.  The memory parameter in EWMA was set within the typical range at $\lambda=0.3$.  BIC was used as the PELTS change point penalty.  We did not include the Bayesian approach in the simulation study due to its comparatively larger computational time.  Section \ref{UWTF}, however, establishes its competitiveness with the hybrid approaches.

We used a logistic regression to summarize this experiment, testing a number of factors for significance.  In both hybrid approaches to model selection, we found that the magnitude of the jumps and a correspondence between the direction of the first derivative at the location of the jump and the jump itself to positively affect detection in a significant way.  Similarly, the derivative, the second derivative, and the complexity of the trend measured by the root mean squared value of the second derivatives, as well as a correspondence between the direction of the second derivative at the location of the jump and the jump itself negatively affect detection in a significant way.  We found that the location of the disturbance relative to the edges is significant in the elbow approach but not the AICc approach.  Additionally, the level of noise in the hybrid approach negatively affects the likelihood of detection.  Noise level is also negatively correlated with with detection percentage in the hybrid elbow approach, but it does not appear as such  in the logistic regression when combined with other factors.  This particular finding is an odd result, but the contribution of the noise to the mean remains less than that of the negative intercept value in the simulated cases.

\begin{table}[H]
\vspace{1em}
\centering
	\caption{Coefficients from logistic regression on simulation study.  Significance measured in p-values: 0 `***', 0.001 `**', 0.01 `*', 0.05 `.' }
	\vspace{1em}
	\begin{tabular}{|c|c|c|}
  	\hline
  	\textbf{Coefficients} & \textbf{Hybrid AICc} & \textbf{Hybrid Elbow} \\
 	 \hline
	intercept & 6.260e-01 & -6.687e-01 ***\\
  	abs(jump.size) & 4.264e+00 *** & 1.427e+00 *** \\
  	abs(derivative) & -2.721e-02 & -2.167e+00 * \\
  	abs(second.derivative) & -3.687e+00 *** & -2.017e+01 *** \\
  	jump.size:derivative & 1.853e-01 *** & 1.203e-02 ***  \\
  	jump.size:second.derivative & -6.629e-01 *** & -2.469e-01 *** \\
  	abs(location - middle) & -4.972e-04 & -5.641e-03 *** \\
  	noise & -2.678e+00 *** & 5.801e+00 *** \\
	RMS.second.derivative & -8.796e+00 *** & -5.600+00 ***\\
  	\hline
	\end{tabular}
\end{table}

%
%

\begin{figure}[H]
    \centering
    \includegraphics[scale=.45]{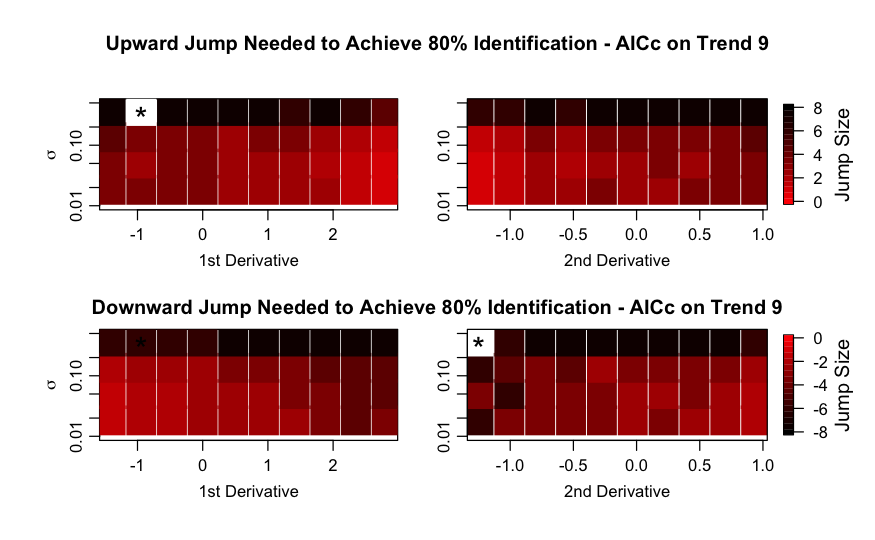}
    \caption{A comparison of the jump size needed to empirically achieve $80\%$ identification of change points over a variety of noise levels across 1st and 2nd derivative values in trend~$9$.  Jumps are split into upward and downward categories. A ``$*$" represents conditions under which a sufficient jump size was not tested to achieve $80\%$ success.}
    \label{jumpsize.trend9}
\end{figure}

We found the effects of the derivative and concavity on each trend to be intuitive yet surprising in the hybrid methods.  Large differences in consecutive measurements result in a greater chance the model identifies the location as a change point.  This is illustrated in the analysis of trend 9 as shown in Figure \ref{jumpsize.trend9}.  We can see that detection is more likely when the derivative of the trend and jump have the same sign.    Additionally, we found that jump detection is more likely when the second derivative of the trend and the jump have opposite signs also seen in Figure \ref{jumpsize.trend9}.    The lack of smoothness in these figures is due to the random sampling of change point locations and specifics of a given trend.

\begin{figure}[H]
    \centering
    \includegraphics[scale=.5]{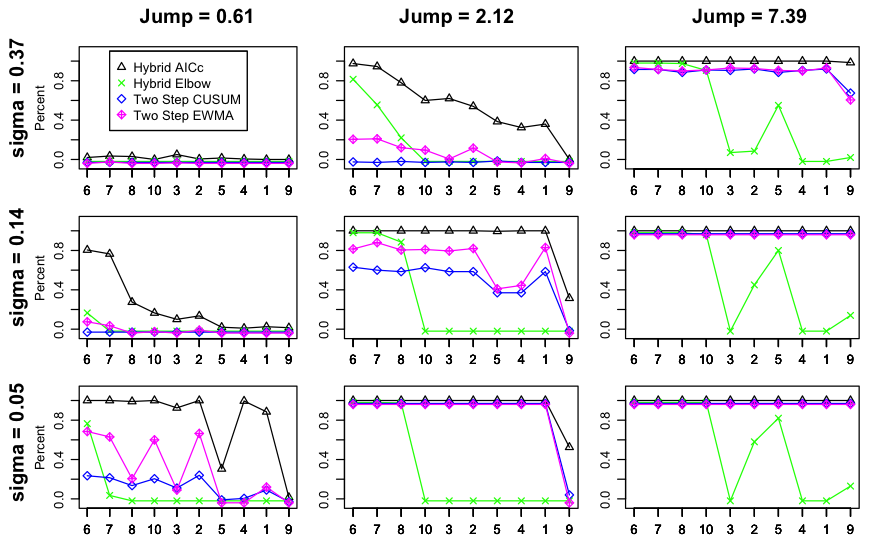}
    \caption{A comparison of methods across jump sizes and noise levels measured by their detection percentage.  The trends along the horizontal axis are ordered from simple to complex via the $L_2$ norm of the second derivative.}
    \label{methods.comparison}
\end{figure}

We compared the hybrid approaches to the two-step approaches using CUSUM, EWMA, and PELTS segmentation.  We found the PELTS approach to be entirely ineffective over the jump sizes we tested, so we limit our comparison to the two hybrid approaches, CUSUM, and EWMA.  In Figure \ref {methods.comparison}, we see a snapshot of the results over a handful of noise levels and anomaly sizes.  We found that all approaches are sensitive to the level of noise in the trend as well as the complexity of the trend, which we measure with the $L_2$ norm of the second derivative.  We found across all results that the hybrid AICc approach did better than the two-step EWMA and CUSUM.  Additionally, we found that at times the hybrid elbow approach is competitive with the two-step approaches but only when the trend is simple. Otherwise it is worse than the two-step EWMA and CUSUM.  In the case of this simulation study, trends 6, 7, and 8 are simple.   

The rate of false positives is also important to mention.  The hybrid elbow and the two-step methods produced close to zero false positives under the tested levels of noise and appeared unaffected by the location of the anomaly.  The hybrid AICc produced few or no false positives in trends with fewer degrees of freedom. However, it is prone to false positive identifications at the boundaries in trends that are highly non-linear at the edges. This is not surprising given the nature of the cubic smoothing spline, which is linear on the boundaries.  Because all false positive identifications using the hybrid AICc approach were located on the ends of the time series, they are easily eliminated by not flagging locations near the beginning and end of each time series.  We found that ignoring $\bgamma$ values in the first and last five values of the time series was sufficient.  Figure \ref{penalty.comp} in the Section \ref{UWTF} illustrates the proclivity of the AICc approach to choose models with fewer degrees of freedom than the other approaches.

\section{Analysis of Urban Water Treatment Facility Dataset} \label{UWTF}

UWTF provided one year of headloss data on sixteen filters, measured at five minute intervals.  The lengths of the filtration cycles ranged from 30 minutes to 48 hours.  Details of the data cleaning and separation of the cycles are found Appendix \ref{separation}.  A number of factors appear to influence cycle lengths including: water quality, weather, season, operator preferences, and maintenance events. Operators typically ended a cycle and initiated a cleaning, referred to as a backwash, when headloss reached 8-8.5 units, but on numerous occasions, cycles were ended after only a short time, long before reaching the 8-8.5 range. The net result of these factors is that the cycles are quite heterogenous, differing in length, concavity, and trend across the different filters, making detection of anomalous cycles more difficult.  

The two methods were applied to 3,225 cycles, of which $65$ anomalous cycles were identified by the facility engineers. The Bayesian approach used four MCMC chains of $1000$ with a burn-in of $200$ and required that disturbances be at least 0.15 units headloss. This value was chosen in consultation with the water engineers at UWTF. They deemed smaller disturbances insignificant.  In the hybrid approach, we attached a stopping limit on the fast iterative shrinkage-thresholding algorithm (FISTA), described in Appendix \ref{fista}, of 1e-4 and applied the same size threshold on disturbances used in the Bayesian approach.  This threshold should reasonably vary across applications.    

\begin{figure} [H]
    \centering
    \includegraphics[scale=.8]{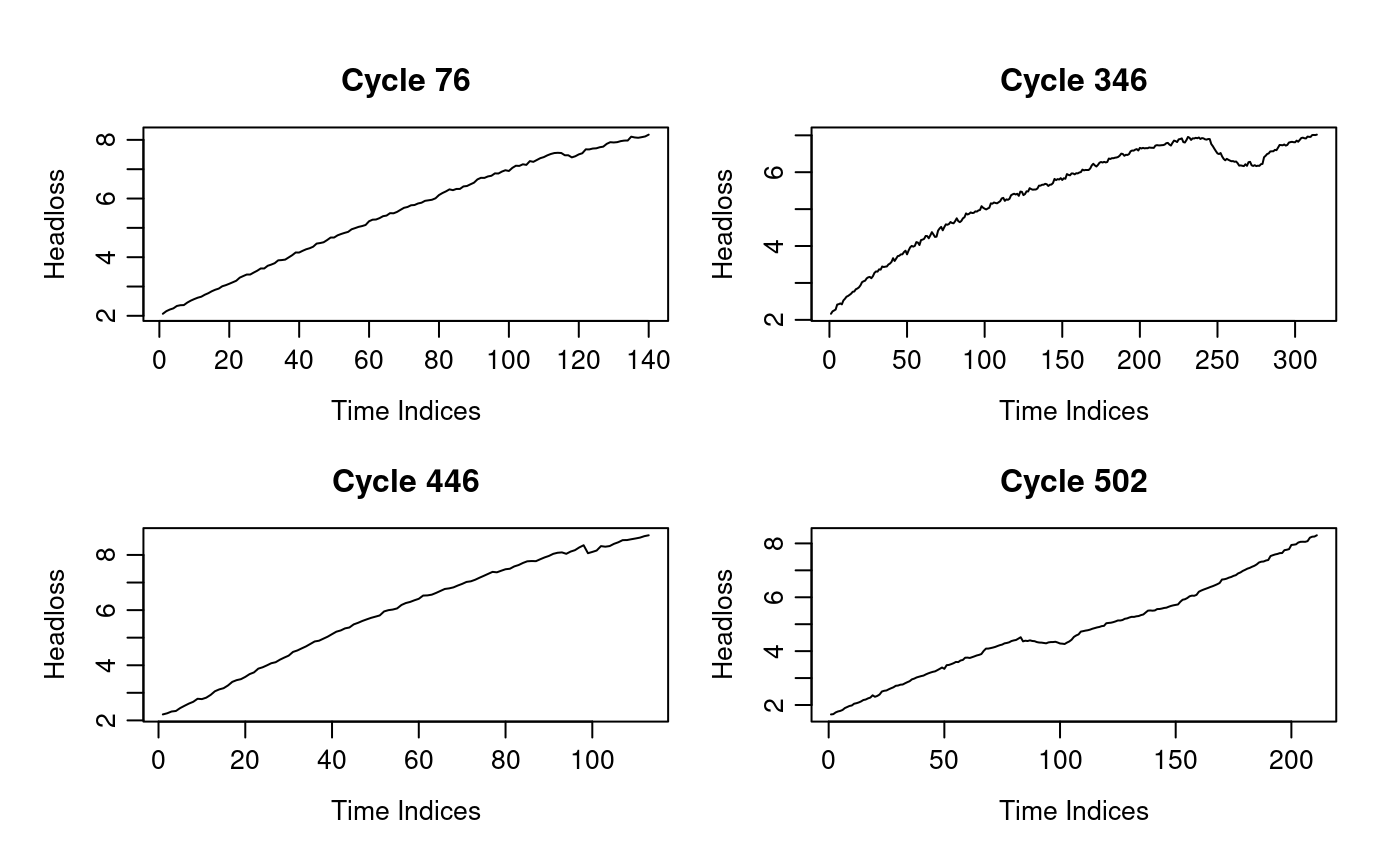}
    \caption{The hybrid elbow and Bayesian approaches missed all four of these cycles.  The hybrid approach missed cycles 76 and 502, but identified cycles 346 and 446.}
    \label{missed}
\end{figure}

The results of the headloss analysis are found in Table \ref{results}.   The Bayesian, the hybrid AICc, and the hybrid elbow approaches were successful in identifying nearly all of the anomalous cycles, each identifying at least 61 of 65 anomalous cycles. See Figure \ref{missed} for images of the four missed cycles.  On each of these cycles, the Bayesian approach did in fact identify disturbances, but at sizes of 0.08, 0.12, 0.11, and 0.12, which are below the 0.15 unit threshold.  One of the two missed cycles by the hybrid AICc approach was also flagged but was beneath the threshold and disregarded, with a disturbance of size 0.11.  

\begin{center}
\begin{table}[H]\label{results}
\centering
\caption{Confusion matrix for approaches in identifying anomalous cycles flagged by the UWTF engineers.}
\textbf{Hybrid AICc/ Hybrid Elbow/ BHM/ CUSUM/ EWMA } \\
\vspace{0.05 in}
\begin{tabular}{ cc|c|c} 
  \multicolumn{4}{c}{PREDICTED}  \\ 
\multirow{5}{*}{\rotatebox{90}{ACTUAL}} & & Anomaly & Normal \\ \cline{2-4}
    & Anomaly & {\textbf{63/61/61/55/47}} & 2/4/4/10/18 \\ \cline{2-4}
    & Normal & 49/63/23/679/305 & \textbf{{3111/3097/3137/2481/2855}} \\ \cline{2-4}
\end{tabular}\\
\end{table}
\end{center}

Cycles 76, 346, and 502 are interesting because they could more aptly be described as a disturbance in slope instead of a change in mean.  Section \ref{extensions} discusses how to adapt the model to search for this type anomaly instead. With this adapted method, each of these cycles is identified by the hybrid AICc approach as anomalous in the first derivative.    

\begin{figure}[H]
    \centering
    \begin{minipage}{.5\textwidth}
      \centering
      \includegraphics[scale=.45]{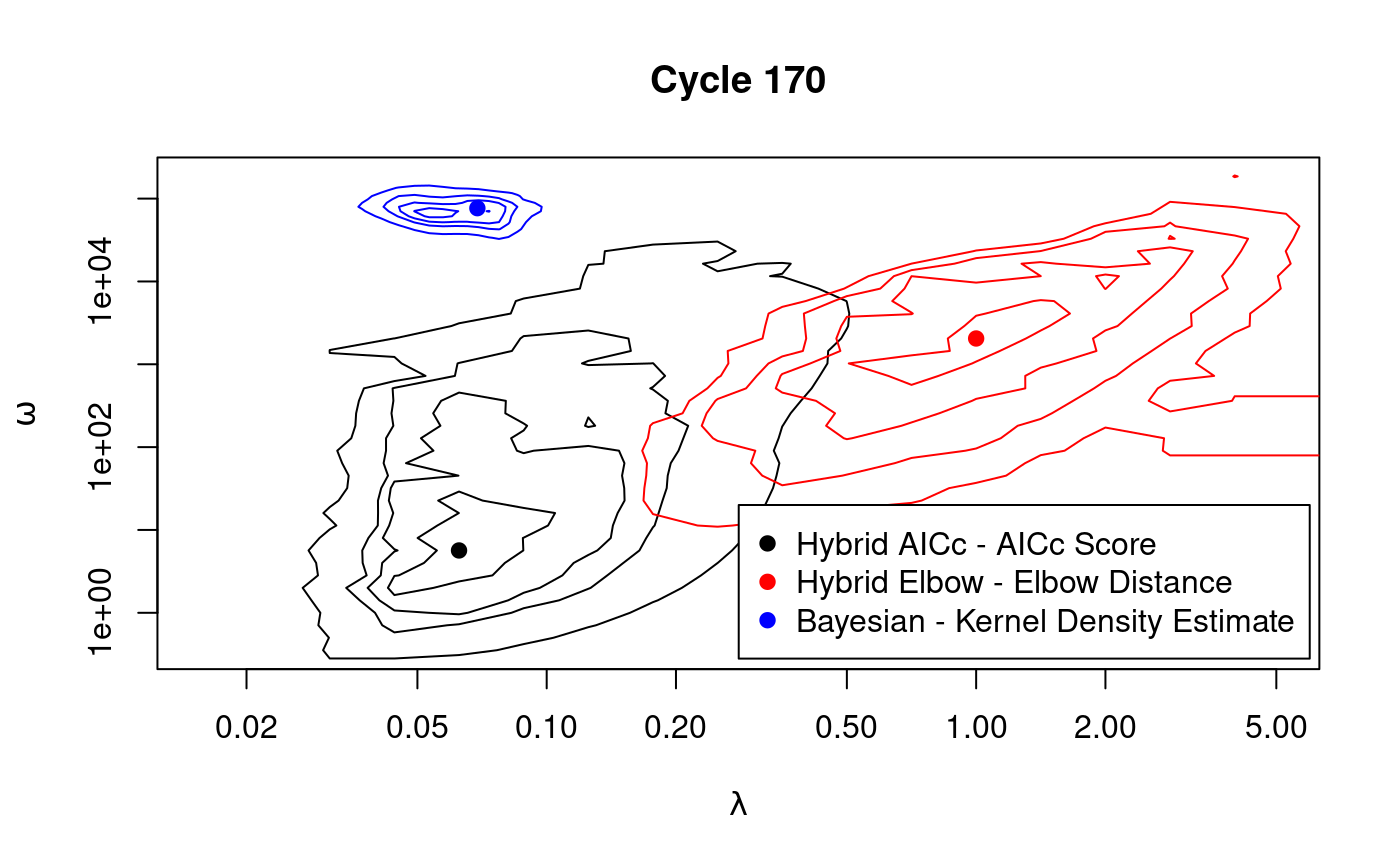}
    \end{minipage}%
    \begin{minipage}{.5\textwidth}
      \centering
      \includegraphics[scale=.45]{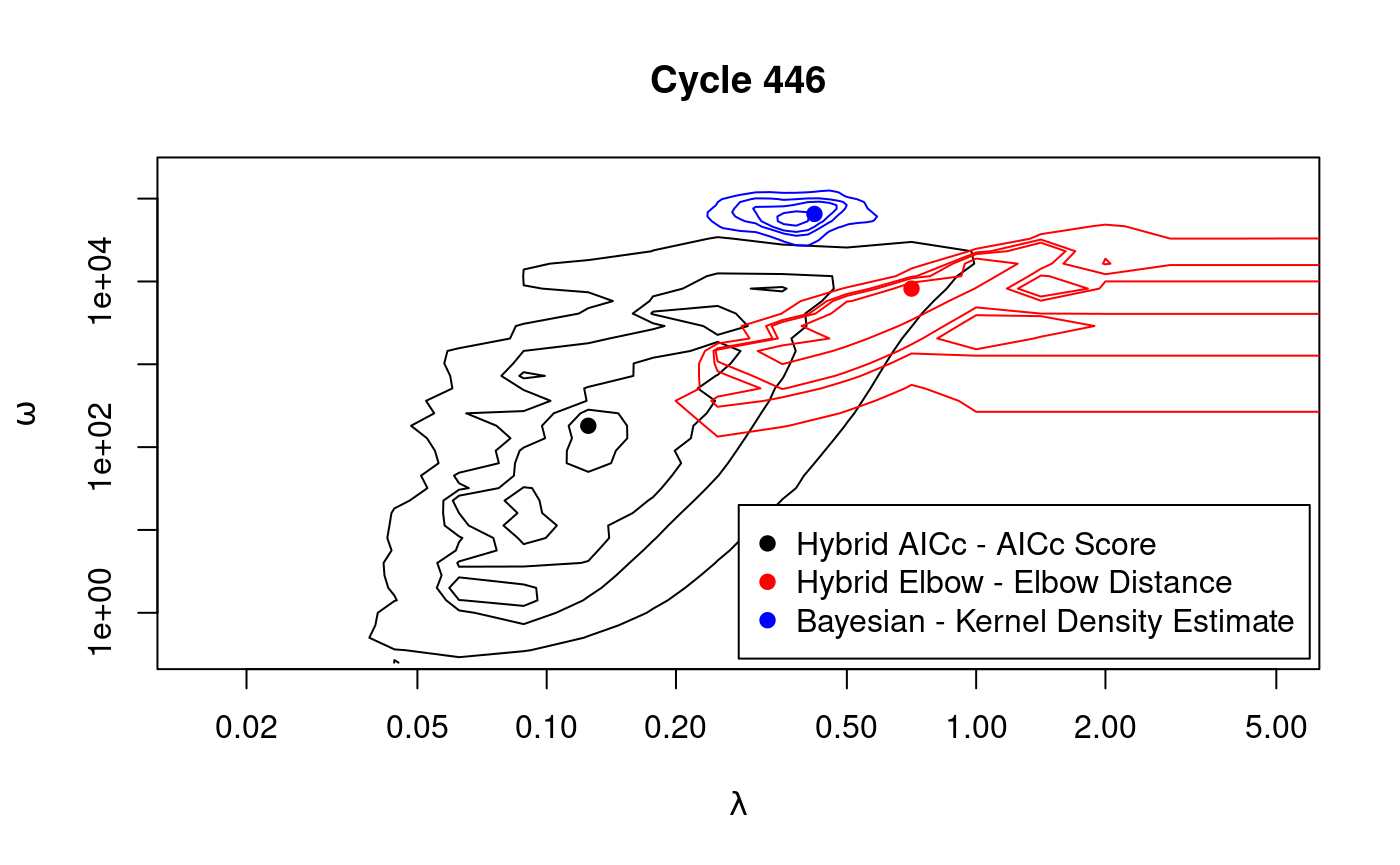}
    \end{minipage}
    \caption{A comparison of optimal penalty parameters, with contour lines at 0.8, 0.9, 0.95, and 0.99 over the three selection metrics, the LASSO penalty on the rough function along the horizontal axis and the smoothing penalty on the trend along the vertical axis.}
    \label{penalty.comp}
\end{figure}

The simulation study demonstrated that the hybrid elbow approach is not as robust to complex trends as the other new approaches. Here, we observe in the UWTF analysis that the elbow is prone to choosing models with a larger LASSO penalty parameter.  In Figure \ref{penalty.comp}, we see a comparison of penalty parameter selection across across the hybrid AICc and elbow approaches and the Bayesian approaches on two cycles.  The first is cycle 170 featured in Sections \ref{General Model} and \ref{bayes}. The second is cycle 446, which was identified by the hybrid AICc approach, missed by the hybrid elbow, and identified but below the threshold by the Bayesian approach.   In all cycles that we checked, we observed the same triangle arrangement of the optimal penalty parameters, the elbow prioritizing simpler rough functions and the Bayesian approach settling on simpler smooth trends.

\begin{figure}[H]\label{zoomin}
    \centering
    \includegraphics[scale=.7]{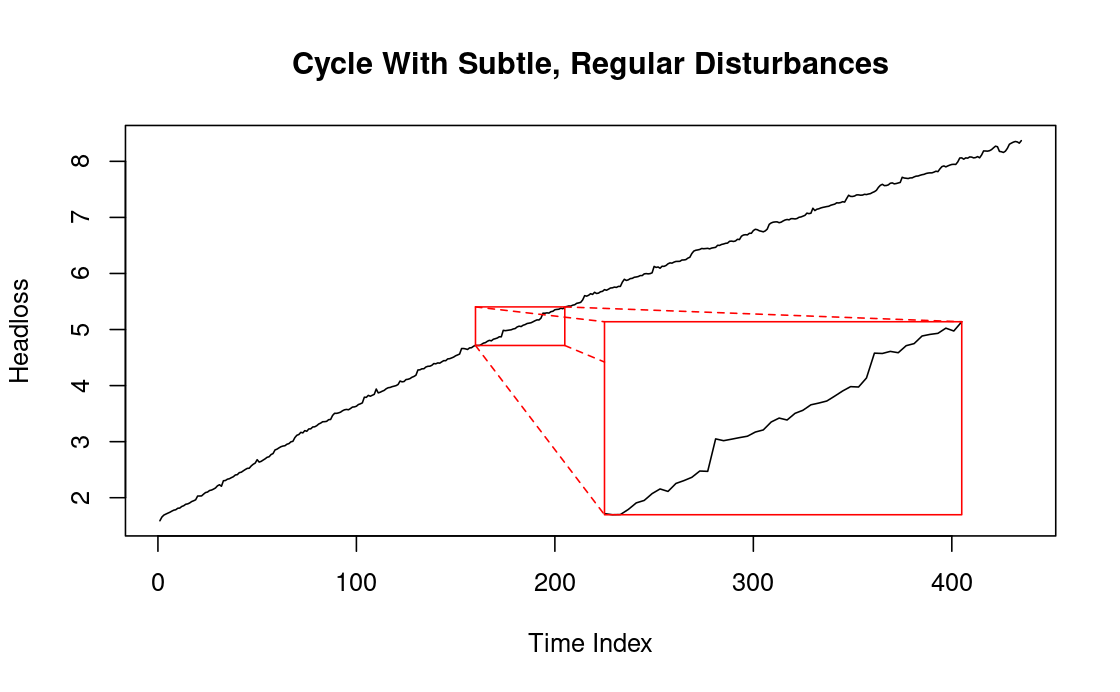}
    \caption{Cycle 3009, one of many cycles identified only by the Bayesian method as anomalous and originally thought to be a false positive. Upon zooming in, however, one can see the irregular, sawtooth nature of the measurements. } 
\end{figure}

Before setting a minimum size on disturbance identification, both the hybrid AICc and BHM approaches yielded a large number of false positives, roughly $400$ of the $3141$ normal cycles.  Upon closer examination however, the models had identified entire months of subtle, odd behavior.  These cycles are in addition to the originally identified $65$ cycles in our key.  One such cycle can be seen in Figure \ref{zoomin}.  Nearly all of these newly identified cycles have a subtle, sawtooth nature to their measurements that is only barely visible.  These anomalies were inspected and determined to not be problematic, but their identification speaks to the effectiveness of this new method even on small discontinuities.

\begin{figure}[H]
    \centering
    \includegraphics[scale=.4]{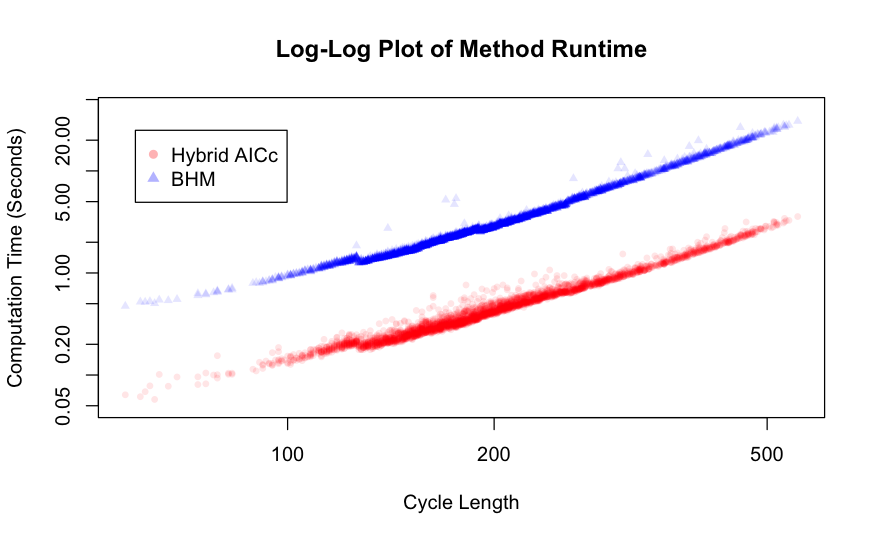}
    \caption{Computation times for anomaly detection.  The Hybrid approach is approximately twenty times faster than the Bayesian approach.} 
    \label{comp time}
\end{figure}

When running on four cores on a MacBook Pro (13-inch, M1, 2020), the computation time required for analyzing cycles is seen in Figure \ref{comp time}.  Overall, the hybrid approach is roughly twenty times faster than the BHM.  Beside length of cycle, the main factor in computation time for the hybrid approach is the grid size of $\lambda$ and $\omega$; for the Bayesian GP approach, the main factor is the length of the MCMC chains.

\section{\textbf{Extension to Discontinuities in the 1st Derivative}} \label{extensions}
In both the simulation study and our analysis of the UWTF headloss cycles, we focused on a specific type of anomaly, namely, a jump in the mean of an unknown trend.  Hybrid smoothing provides significant flexibility in identifying what acceptable behavior looks like, even as more simple trends are prioritized.  No form is assumed because the cubic smoothing spline is a nonparametric model.  The addition of the rough function creates a semiparametric model, and it seems reasonable to add another function rough in the first derivative capable of identifying abrupt changes there as well.  By defining a new set of rough basis functions, $\Omega = \Psi^2$, and a new model fit to values of $\bc$, $\bgamma$, and $\mathbf{\kappa}$ is

\[ \by=\Phi \bc + \Psi \bgamma + \Omega \mathbf{\kappa} + \epsilon \]

\noindent By applying the same penalties on $\bc$ and $\bgamma$ and incorporating $\mathbf{\kappa}$ into the LASSO penalty, we have the potential to identify disturbances in the mean and the first derivative.  

\begin{figure}[H]
    \centering
    \includegraphics[scale=.8]{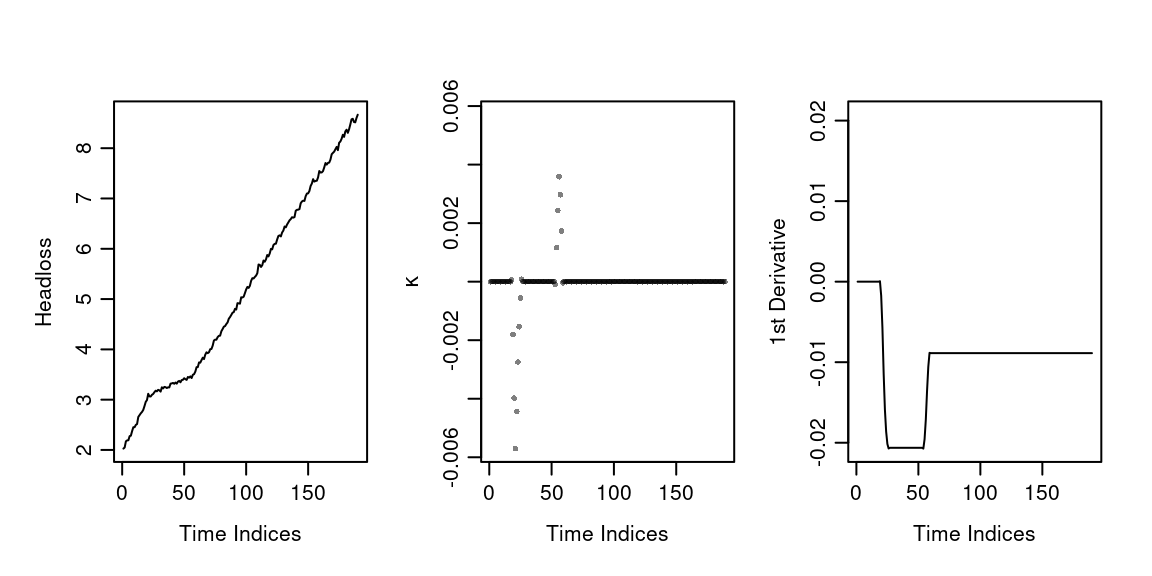}
    \caption{Successful identification of change points in the first derivative of headloss cycle 1102 from UWTF.  The left plot is the headloss cycle.  The middle plot contains the coefficients of the rough 1st derivative basis functions. The right plot is the reconstruction of the first derivative of the combined rough function.} 
    \label{2nd.model}
\end{figure}

This adjusted model can be fit in the same manner as the original.  Figure \ref{2nd.model} is an example of the new model successfully identifying a change in the first derivative.  We believe this additional approach is worth further study, however, we only briefly discuss it here.   

\begin{figure}[H]
    \centering
    \includegraphics[scale=.8]{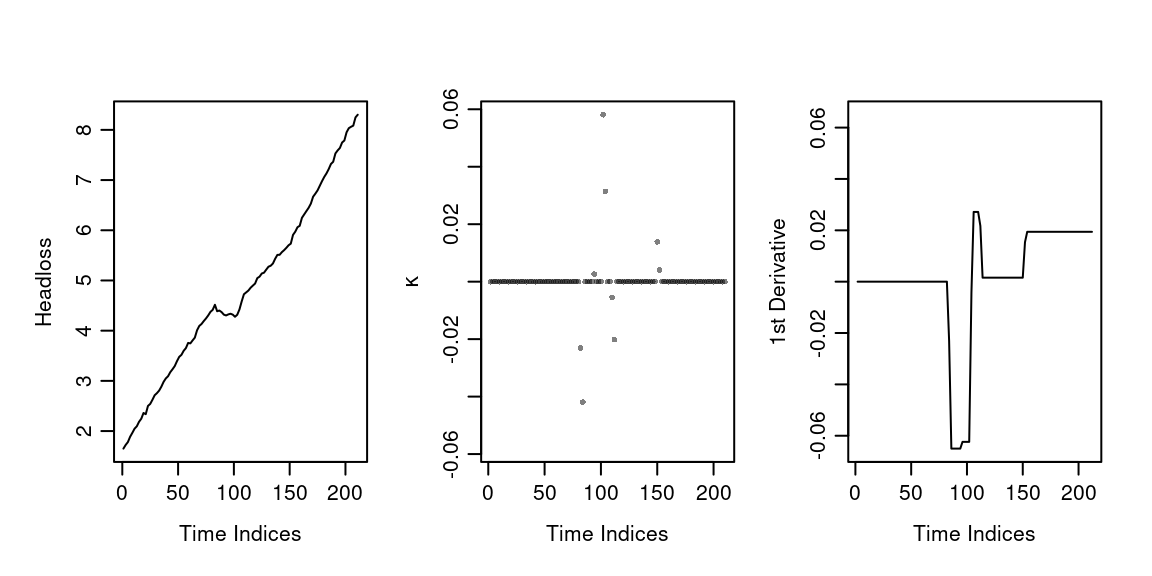}
    \caption{Successful identification of change points in headloss cycle 502 from UWTF once the time series is thinned.  The left plot is the headloss cycle.  The middle plot contains the coefficients of the rough 1st derivative basis functions. The right plot is the reconstruction of the first derivative of the combined rough function.} 
    \label{missed.then.found}
\end{figure}

Our limited investigation of this approach has shown it to be quite sensitive to the complexity of a trend.  To increase the robustness of this method, we found that thinning the time series to simplify the trend was often an effective tool when paired with the hybrid AICc approach.  We removed every other data point and reran the three UWTF cycles missed by the hybrid AICc approach. The hybrid AICc approach then identified two anomalous cycles (76 and 502) that it had originally missed when searching for a disturbance in the linear trend instead of a change in mean. Figure \ref{missed.then.found}shows cycle 502, one of the missed UWTF cycles.    Cycle 346 is also identified as a change in linear trend using with the hybrid AICc approach.

\section{\textbf{Conclusion}}\label{Conclusions}
 
Recovering a smooth trend from a time series using quadratic roughness penalties, an $L_2$ type criteria, or selecting a subset of variables using the LASSO, an $L_1$ type criteria, are both well established and successful data analysis tools. Here, we have combined these models to detect anomalous signals in process monitoring. At the outset, this may not have seemed possible due to the ambiguity of separating a signal into smooth and anomalous components. However, we found that for our motivating application in water treatment, the method works well, showing false negative rates of a few percent and low false positive rates when tuned to remove very small disturbances. Additionally, both the BHM and the hybrid AICc approaches provided an opportunity to detect more subtle discontinuities in the filtering cycles that are not easily spotted by visual inspection.   

In our simulation study, we observed that the hybrid AICc approach outperforms two-step CUSUM, EWMA, and segmentation approaches for detecting changes in the mean of data in all tested cases.  Additionally, the hybrid elbow approach appears useful in detecting change points in simple trends.  We saw that the selection of penalty parameters plays an important role in the effectiveness of each method.

Besides having success for a substantial application, this work also provides four methodological contributions. One is the development of a method for detecting anomalies in trended data in a single step.  The second is testing in a Monte Carlo setting the simplex strategy for choosing the two smoothing/sparsity  parameters in the frequentist method. This parameter choice drawn from machine learning worked surprisingly well and is also fast. Its success motivates some additional investigation into deeper statistical principles as to why this works. The third contribution is in transforming the hybrid model into equivalent forms that make it efficient to estimate and perform an MCMC computation. This includes reducing the minimization for  the hybrid estimator to a linearly transformed problem that just has the $L_1$ minimization. Finally, for computing the BHM, we were able to recast the model  to avoid more expensive MH updates and also to reparametrize the linear models to accelerate mixing. The net result is an efficient MCMC sampler for a complex hierarchical model.

One current topic in spatial statistics is models for non-Gaussian fields. We note that the BHM is not tied to a particular GP, and one could add extra parameters for the covariances typically used in geostatistics. The  $L_2$/$L_1$ model would make  the estimated fields more interpretable given the separation into a GP and non-Gaussian process components. One strategy for handling the analysis of large spatial data sets is to expand the spatial fields in basis functions with a multivariate normal prior on the coefficients. This is similar to the $L_2$ part of our hybrid smoother, and it is reasonable to extend this to additional basis functions to capture discontinuities.  This additional part of the model could capture transitions between land/water boundaries for climate data or demographic and municipal regions when one considers social data.  In general, this hybrid model has potential to go beyond representations of data that are not supported by just a smooth surface and random noise.

\section{\textbf{Acknowledgements}}
The authors would like to thank Mike Wakin and Mevin Hooten for useful conversations.

\section{\textbf{Funding}}
This research is supported in part by the National Science Foundation under NSF Award 1924146
HDR DSC: Collaborative Research: Modernizing Water and Wastewater Treatment through Data Science Education \& Research (MoWaTER).

\bibliographystyle{Chicago}
\bibliographystyle{chicago.sty}
\bibliography{Bibliography-MM-MC}

\appendix

\section{Cycle Separation}\label{separation}
\begin{figure}[H]
    \centering
    \begin{minipage}{.5\textwidth}
      \centering
      \includegraphics[scale=.12]{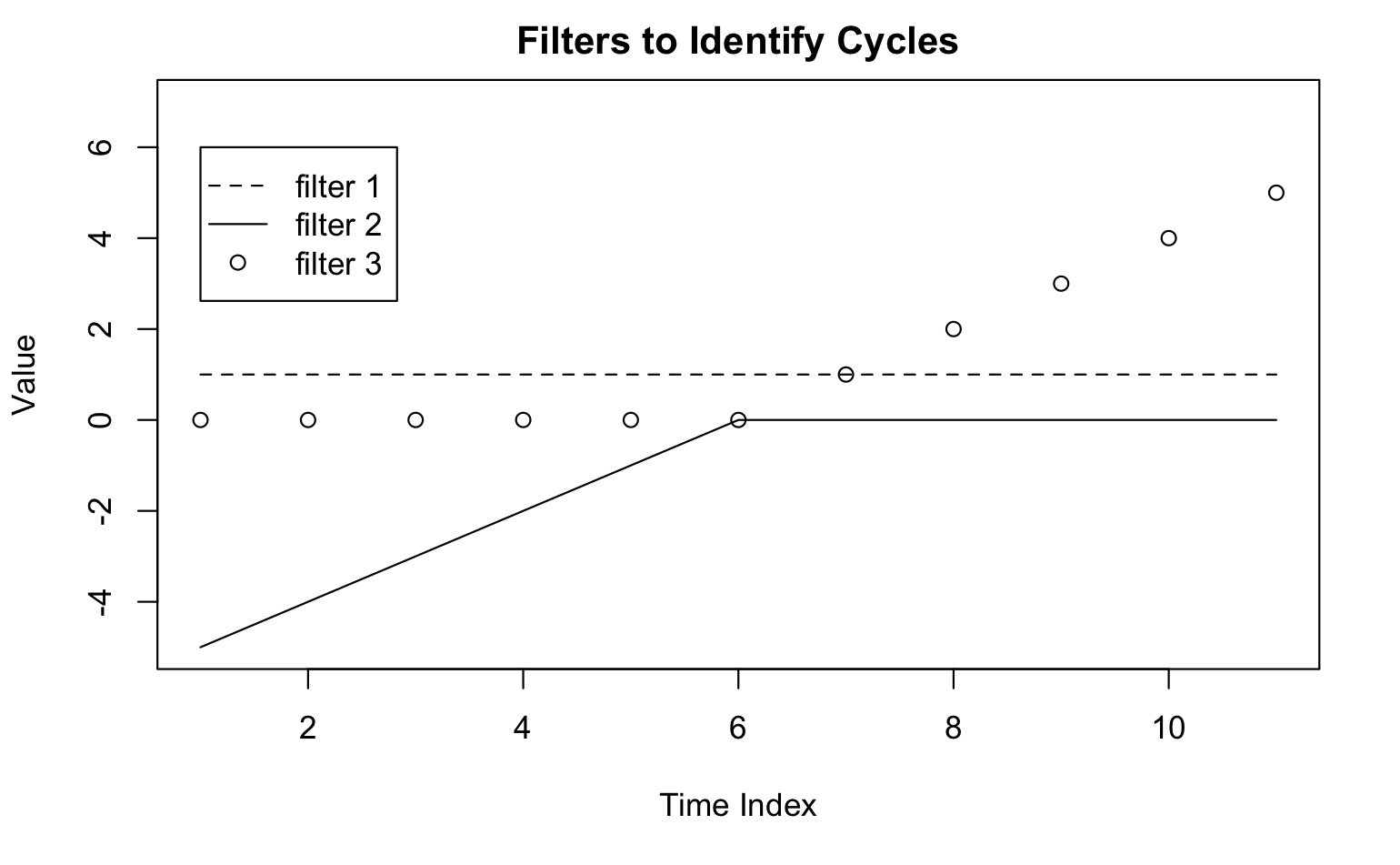}
    \end{minipage}%
    \begin{minipage}{.5\textwidth}
      \centering
      \includegraphics[scale=.12]{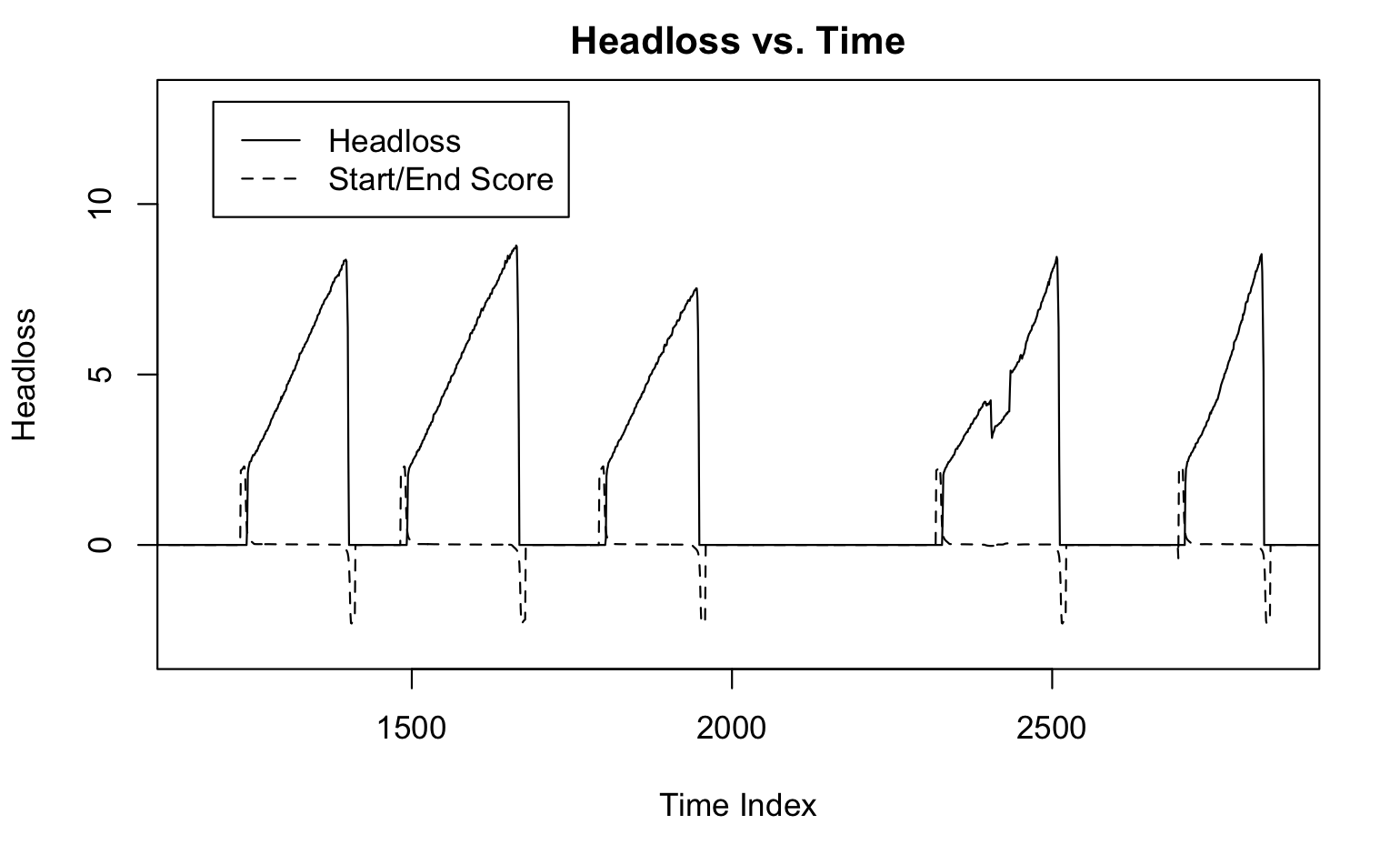}
    \end{minipage}
    \caption{Left: The three filters passed over the headloss data to identify the start and end points of each cycle.  Right: The dotted lines show the scores of the filtering operation used to determine the start and end of each cycle.}
    \label{filter}
\end{figure}

The headloss data from UWTF were not reported as individual cycles, so the first step in the analysis process was separating the cycles. There are some approaches to separating the headloss data into individual cycles that can be used.  Ideally, headloss data are accompanied by a measurement of whether the filter is on or off, which makes the separation process simple.  If the on/off information is not present, then cycles can be separated using a filtering operation on the data.  The one used in this paper can be seen in Figure \ref{filter}. Once the inner products of the filter and data are computed at each point in the cycle, these values can be combined in various ways to identify the start and end points of the cycles as seen in the figure.  The start and end points identified using this approach are a near perfect match of the on/off data reported by UWTF; however, for a number of cycles, the on/off data was found to be inaccurate and needed to be corrected.

\section{Solving the LASSO}\label{fista}
The optimization problem in (\ref{reducedLASSO}) can be solved using a standard LASSO implementation after transforming $\by$ and $\Psi$ by multiplication by $\bW(\omega)$.  We adopt the fast iterative shrinkage-thresholding algorithm (FISTA) algorithm proposed by \citet{beck2009fast}.

First, we set $\by^* = \bW(\omega)\by$ and $\bX= \bW(\omega) \Psi$.  The steps of the FISTA algorithm are as follows:

\begin{itemize}
    \item Step 0: Choose an initial value for $\bgamma$.  Here, $s_0=1$ and $q_0=\bgamma_0$.
    \item Step 1: Apply gradient decent on $\bgamma$ $$\bz_k= \bgamma_k - \tau \bX^T(\bX\bgamma_k - \by^*)$$
    where $\tau$  is the reciprocal of the largest eigenvalue of $\bX^T \bX$.  See Appendix \ref{lipschitz}. \\
    \item Step 2: Perform soft thresholding on $\bz_k$
    $$q_{k}=(|\bz_k|-\alpha)_+\text{sign}(\bz_k) $$ where $\alpha = \frac{\lambda \tau}{2}$
    \item Step 3: Update $s_k$, as.
    $$s_{k+1}=\frac{1+\sqrt{1+4s_k^2}}{2}$$ 
    \item Step 4: Update $\bgamma$, as.
    $$\bgamma_{k+1}=q_{k}+\frac{s_k-1}{s_{k+1}}(q_k-q_{k-1})$$
    \item Step 5: Repeat steps 1 through 4 until convergence.
\end{itemize}
\noindent

\section{Gibbs Updates and Posterior Distributions}\label{posteriors}
 All of the full conditional distributions for the OBHM are  straightforward to derive with the possible exception of $\tau_j^2$, which is derived in Appendix \ref{taupost}.  Below is a list of the full conditional posterior distributions of each level.  Note that in the posteriors of $\lambda^2$ and $\omega$, the second parameter is the rate.

\begin{align}
    \bbeta^*|\cdot &\sim N_n(A^{-1}b,A^{-1}) \text{  where  } A=\frac{X^TX}{\sigma^2} \text{  and  } b=\frac{X^T(Y-\Psi^*\bgamma^*-H\bg)}{\sigma^2}\\
    \bgamma^* |\cdot & \sim N_n(A^{-1}b,A^{-1}) \text{  where  } A=\frac{\Psi^{*T}\Psi^*+F^{-1}}{\sigma^2} \text{  and  } b=\frac{\Psi^{*T}(y-X\bbeta^*-H\bg) + F^{-1}J\bg}{\sigma^2} \\
    \sigma^2|\cdot & \sim \text{inv-gamma}\left(\frac{2n-1}{2}, r\right)\\
    \text{where } r&=\frac{(y-X\bbeta^*-\Psi^*\bgamma^*-H\bg)^T(y-X\bbeta^*-\Psi^*\bgamma^*-H\bg)}{2} + \frac{(\bgamma^*-J\bg)^T F^{-1} (\bgamma^*-J\bg)}{2} \nonumber  \\ 
 \lambda^2|\cdot &\sim \text{gamma
    }(\alpha_{\lambda^2}+n-1, \beta_{\lambda^2} + \sum_{j=1}^n \tau_j^2/2) \\
    \frac{1}{\tau_j^2}|\cdot &\sim \text{inv-Gaussian}\left(\sqrt{\frac{\sigma^2\lambda^2}{(\bgamma^*-J\bg)_j^2}},\lambda^2\right) \\
    \bg|\cdot &\sim N_n(A^{-1}b,A^{-1}) \text{  where  } A=\omega K^{-1} + \frac{H^TH+J^TF^{-1}J}{\sigma^2}\\
    &\text{and  } b=\frac{H^T(Y-X\bbeta^*-\Psi^*\bgamma^*)+J^TF^{-1}\bgamma^*}{\sigma^2} \nonumber \\
    \omega &\sim \text{gamma}\left(\frac{n}{2} + \alpha_{\omega}, \frac{\bg^TK^{-1}\bg}{2}+\beta_\omega\right)
\end{align}

Upon completion of the MCMC, it is necessary to parametrize back to $\bbeta$ and $\bgamma$.  This is done by finding $\bgamma=\bgamma^*-Jf$ and $\bbeta =\bbeta^*-(X^TX)^{-1}X^T(\Psi \bgamma+\bg)$. Although the draws were made of $\bgamma^*$ and $\bbeta^*$, the inferences can still be made on $\bgamma$ and $\bbeta$.

\section{Other Constructions of $\Psi$} \label{ConstructPsi}
The piecewise stochastic portion of the model, $g(t)=\Psi\bgamma$, designed to detect disturbances in the mean holds some intriguing possibilities if we allow for other formulations of $\Psi$. In the recovery of $g(t)$, we found that we can dramatically reduce the error bounds by reformulating $\Psi$ to reconstruct $g(t)$ from the center of the time series instead of the beginning.  $\Psi$ is an upper triangular matrix of ones on the left half of the matrix and a lower triangular matrix of ones on the right side of the matrix, and the middle column is removed from the matrix because the mean is already identified elsewhere in the model.  This reduces the size of the largest credible intervals around any location in $g(t)$ by a factor of two.  This is the construction of $\Psi$ that we used in the implementation of all three approaches.  An important note, however, is that the choice of $\Psi$ does not change the detection of significant $\bgamma$ coefficients or their distributions.

\section{The Lipschitz Constant in FISTA} \label{lipschitz}
Convergence of the FISTA algorithm requires that $h:\mathcal{R}^n\rightarrow  \mathcal{R}$ is convex and differentiable and that $\nabla h$ is Lipschitz continuous with constant $L\geq 0$.  This means that $$\|\nabla h(x) - \nabla h(y)\| \leq L \|x-y\|$$ for all $x$ and $y$.  $L$ can be taken to its limit, and the Lipschitz constant becomes the largest eigenvalue of the Hessian matrix of $h$.  In both programs $f = \|\bW(\omega)\by-\bW(\omega) \bgamma\|^2_2$, so $\nabla^2h =\Psi^T\bW(\omega)^T\bW(\omega)\Psi =2\bX^T\bX$.

\section{Proof Model Has Functionally Laplace Prior} \label{LaplaceIdentity}
Here, we demonstrate that the proposed hierarchical model is the functional equivalent of the double exponential requires integrating out each of the $\tau_j$'s.
\begin{align}
[\bgamma|\sigma^2,\lambda^2] &\propto  \prod_{j=1}^n \int_0^{\infty} \frac{1}{\sqrt{\sigma^2\tau_j^2}}exp(-\frac{1}{\sigma^2\tau_j^2}\bgamma_j^2)\frac{\lambda^2}{2}exp(-\frac{\lambda^2\tau_j^2}{2})d\tau_j^2\\
&\propto  \prod_{j=1}^n \int_0^{\infty} \frac{1}{\sqrt{\sigma^2\tau_j^2}}exp(-\frac{1}{\sigma^2\tau_j^2}\bgamma_j^2)\frac{\lambda^2}{2\sigma^2}exp(-\frac{\lambda^2\tau_j^2\sigma^2}{2\sigma^2})\sigma^2 d\tau_j^2
\end{align} 
using the change of variables $s=\sigma^2\tau_j^2$, $a=\sqrt{\frac{\lambda^2}{\sigma^2}}$, we get 
\begin{align}
    [\bgamma|\sigma^2,\lambda^2] &= \prod_{j=1}^n \int_0^\infty \frac{1}{\sqrt{2\pi s}} e^{-\bgamma^2/(2s)} \frac{a^2}{2} e^{-a^2s/2}ds
\end{align}
And by Identity (\ref{laplace}),
\begin{align}
    [\bgamma|\sigma^2,\lambda^2] &= \frac{a}{2}e^{-a|\bgamma|}\\
\end{align}

\section{Conditional Posterior of $\tau_j^2$ } \label{taupost}
The posterior of $\frac{1}{\tau_j^2}$ is the inverse Gaussian distribution discussed by \citet{chhikara1989inverse}.  Here is an explanation of the formulation found in \citet{park2008bayesian}.\\

\begin{align}
    [\tau_j^2] &\propto \frac{1}{|\sigma^2F|^{1/2}}exp(\frac{-1}{2\sigma^2}\bgamma^TF^{-1}\bgamma)\prod_{j=1}^n \frac{\lambda^2}{2}e^{-\lambda^2\tau_j^2/2} \\
    & \propto \prod_{j=1}^n \frac{\lambda^2}{2(\sigma^2\tau_j^2)^{1/2}}exp(-\frac{\lambda^2\tau_j^2}{2} - \frac{1}{2\sigma^2}\frac{\bgamma_j^2}{\tau^2_j})
\end{align}
Applying the change of variables $x=\frac{1}{\tau^2_j}$ results in 
\begin{align}
    [x] &\propto \prod_{j=1}^n x^{1/2} exp(-\frac{\bgamma_j^2}{2\sigma^2}(x+\frac{\lambda^2\sigma^2}{\bgamma_j^2x}))x^{-2}\\
    &\propto \prod_{j=1}^n x^{-3/2} exp(-\frac{\bgamma_j^2\lambda^2}{2\sigma^2\lambda^2x}(x^2+\frac{\lambda^2\sigma^2}{\bgamma_j^2}))\\
    &\propto \prod_{j=1}^n x^{-3/2} exp(-\frac{\bgamma_j^2\lambda^2}{2\sigma^2\lambda^2x}(x-\sqrt{\frac{\lambda^2\sigma^2}{\bgamma_j^2}})^2)
\end{align}
Letting $\mu'=\sqrt{\frac{\lambda^2\sigma^2}{\bgamma_j^2}}$ and $\lambda'=\lambda^2$, we get
\begin{align}
     [x]&\propto \prod_{j=1}^n x^{-3/2} exp(-\frac{\lambda'}{2(\mu')^2x}(x-\mu')^2)\\
     \frac{1}{\tau_j^2} &\sim \text{Inv-Gaussian}(\mu',\lambda')
\end{align}

\end{document}